\documentclass[twocolumn,showkeys,nofootinbib]{revtex4}
\usepackage{epsfig}
\usepackage{graphicx}
\usepackage{soul}
\usepackage{color}
\usepackage{amsmath}
\usepackage{stackrel}
\begin{document}
\title{Chimera states in networks of locally and non-locally coupled SQUIDs}
\author{J. Hizanidis, N. Lazarides, G. P. Tsironis}
\affiliation{Department of Physics, University of Crete,
 P. O. Box 2208, 71003 Heraklion, Greece}
\date{\today}
\begin{abstract}
Planar and linear arrays of SQUIDs (superconducting quantum interference devices), 
operate as nonlinear magnetic metamaterials in microwaves. Such {\em SQUID 
metamaterials} are paradigmatic systems that serve as a test-bed for simulating 
several nonlinear dynamics phenomena. SQUIDs are highly nonlinear oscillators 
which are coupled together through magnetic dipole-dipole forces due to their 
mutual inductance; that coupling falls-off approximately as the inverse cube of 
their distance, i.~e., it is non-local. However, it can be approximated by a local 
(nearest-neighbor) coupling which in many cases suffices for capturing the 
essentials of the dynamics of SQUID metamaterials. For either type of coupling, 
it is numerically demonstrated that chimera states as well as other spatially 
non-uniform states can be generated in SQUID metamaterials under time-dependent 
applied magnetic flux for appropriately chosen initial conditions. The mechanism 
for the emergence of these states is discussed in terms of the multistability 
property of the individual SQUIDs around their resonance frequency and the attractor 
crowding effect in systems of coupled nonlinear oscillators. Interestingly, 
generation and control of chimera states in SQUID metamaterials can be achieved 
in the presence of a constant (dc) flux gradient with the SQUID metamaterial 
initially at rest.
\end{abstract}
\keywords{SQUID metamaterials, magnetic metamaterials, chimera states, attractor 
 crowding, synchronization-desynchronization transition}
\maketitle
\section{Introduction}
The notion of metamaterials refers to artificially structured media designed 
to achieve properties not available in natural materials. Originally they were
comprising subwavelength resonant elements, such as the celebrated split-ring
resonator (SRR). The latter, in its simplest version, is just a highly conducting
metallic ring with a slit, that can be regarded as an effectively resistive - 
inductive - capacitive ($RLC$) electrical circuit. There has been a tremendous 
amount of activity in the field of metamaterials the last two decades, the 
results of which have been summarized in a number of review articles 
\cite{Smith2004,Linden2006,Padilla2006,Shalaev2007,Litchinitser2008,Soukoulis2011,
YLiu2011,Simovski2012} and books \cite{Engheta2006,Pendry2007,Ramakrishna2009,
TJCui2010,Cai2010,Solymar2009,Noginov2012,Tong2018}. One of metamaterial's most 
remarkable properties is that of the {\em negative refraction index}, which 
results from simultaneously negative dielectric permittivity and diamagnetic 
permeability. 

\begin{figure}[h!]
\begin{center}
\includegraphics[width=8.5cm]{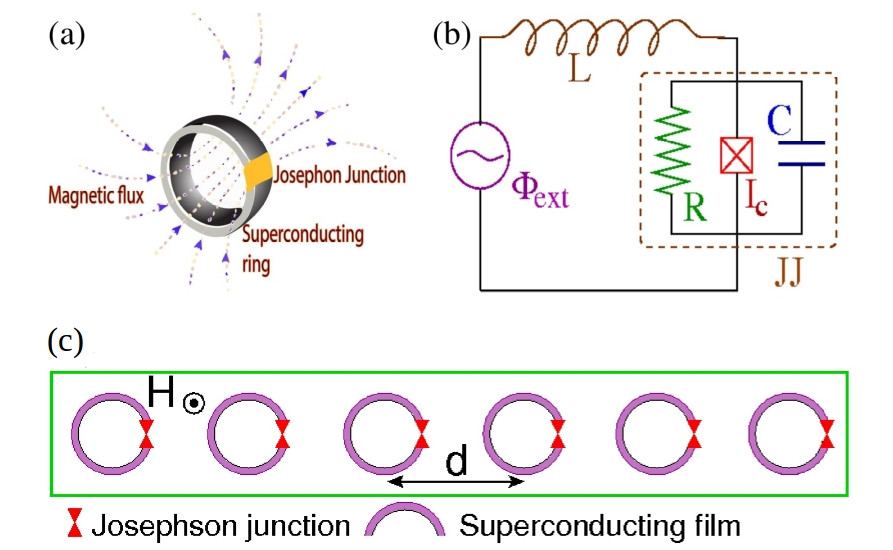} 
\end{center}
\caption{(a) Schematic of a SQUID (superconducting quantum interference device)
 in a magnetic field. (b) Equivalent electrical circuit. 
(c) Schematic top view of a one-dimensional periodic
 array of SQUIDs in a magnetic field $\bf H$}
\label{fig01}
\end{figure}

An important subclass of metamaterials is that of superconducting ones 
\cite{Anlage2011,Jung2014}, in which the elementary units (i.~e., the SRRs) are 
made by a superconducting material, typically Niobium ($Nb$) \cite{Jin2010} or 
Niobium Nitride ($NbN$) \cite{Zhang2012}, as well as perovskite superconductors
such as yttrium barium copper oxide ($YBCO$) \cite{Gu2010}. In superconductors,
the dc resistance vanishes below a critical temperature $T_c$; thus, below $T_c$, 
superconducting metamaterials have the advantage of ultra-low losses, a highly 
desirable feature for prospective applications. Moreover, when they are in the 
superconducting state, these metamaterials exhibit extreme sensitivity in external 
stimuli such as the temperature and magnetic fields, which makes their thermal 
and magnetic tunability possible \cite{XZhang2014}. Going a step beyond, the 
superconducting SRRs can be replaced by SQUIDs \cite{Du2006,Lazarides2007}, 
where the acronym stands for {\em Superconducting QUantum Interference Devices}. 
The simplest version of such a device consists of a superconducting ring 
interrupted by a Josephson junction (JJ) \cite{Josephson1962}, as shown 
schematically in Fig. \ref{fig01}(a); the most common type of a JJ is formed 
whenever two superconductors are separated by a thin insulating layer 
(superconductor / insulator / superconductor JJ). The current through the 
insulating layer and the voltage across the JJ are then determined by the 
celebrated Josephson relations. Through these relations, the JJ provides a strong 
and well-studied nonlinearity to the SQUID, which makes the latter a unique 
nonlinear oscillator that can be actually manipulated through multiple external 
means. 

SQUID metamaterials are extended systems containing a large number of SQUIDs
arranged in various configurations which, from the dynamical systems point of 
view, can be viewed theoretically as an assembly of weakly coupled nonlinear 
oscillators that inherit the flexibility of their constituting elements (i.e, 
the SQUIDs). They present a {\it nonlinear dynamics laboratory} in which numerous 
classical as well as quantum complex spatio-temporal phenomena can be explored. 
Recent experiments on SQUID metamaterials have revealed several extraordinary 
properties such as negative permeability \cite{Butz2013a}, broad-band tunability 
\cite{Butz2013a,Trepanier2013}, self-induced broad-band transparency 
\cite{Zhang2015}, dynamic multistability and switching \cite{Jung2014b}, as well 
as coherent oscillations \cite{Trepanier2017}. Moreover, nonlinear effects such 
as localization of the discrete breather type \cite{Lazarides2008a} and nonlinear 
band-opening (nonlinear transmission) \cite{Tsironis2014b}, as well as 
the emergence of counter-intuitive dynamic states referred to as {\em chimera 
states} in current literature \cite{Lazarides2015b,Hizanidis2016a,Hizanidis2016b}, 
have been demonstrated numerically in SQUID metamaterial models 
\cite{Lazarides2018b}. 

The chimera states, in particular, which were first discovered in rings of 
non-locally and symmetrically coupled identical phase oscillators 
\cite{Kuramoto2002}, have been reviewed thoroughly in recent articles 
\cite{Panaggio2015,Scholl2016,Yao2016}, are characterized by the coexistence of 
synchronous and asynchronous clusters of oscillators; their discovery was 
followed by intense theoretical
\cite{Abrams2004,Kuramoto2006,Omelchenko2008,Abrams2008,Pikovsky2008b,Ott2009,
Martens2010,Omelchenko2011,Yao2013,Omelchenko2013,Hizanidis2014,Zakharova2014,
Bountis2014,Yeldesbay2014,Haugland2015,Bera2016,Shena2017,Sawicki2017,Ghosh2018,
Shepelev2018,Banerjee2018}
and experimental 
\cite{Tinsley2012,Hagerstrom2012,Wickra2013,Nkomo2013,Martens2013,Schonleber2014,
Viktorov2014,Rosin2014,Schmidt2014b,Gambuzza2014,Kapitaniak2014,Larger2015,
Hart2016,English2017,Totz2018}
activities, in which chimera states have been observed experimentally or 
demonstrated numerically in a huge variety of physical and chemical systems.

Here, the possibility for generating chimera states in SQUID metamaterials driven
by a time-dependent magnetic flux by proper initialization or by the application
of a dc flux gradient is demonstrated numerically. The SQUIDs in such a 
metamaterial are coupled together through magnetic dipole-dipole forces due to 
their mutual inductance. This kind of coupling between SQUIDs falls-off 
approximately as the inverse cube of their center-to-center distance, and thus 
it is clearly non-local. However, due to the magnetic nature of the coupling, 
its strength is weak \cite{Trepanier2013,Trepanier2017}, and thus a 
nearest-neighbor coupling approach (i.~e., a local coupling approach) is often 
sufficient in capturing the essentials of the dynamics of SQUID metamaterials. 
Chimera states emerge in SQUID metamaterials with either non-local 
\cite{Lazarides2015b,Hizanidis2016b} or local \cite{Hizanidis2016a} coupling 
between SQUIDs. They can be generated from a large variety of initial conditions, 
and they are characterized using well-established measures. It is also 
demonstrated that chimera states emerge in SQUID metamaterials with zero initial 
conditions using a dc flux gradient; in that case, control over the obtained 
chimera states can be achieved.   

In the next section, a model for a single SQUID that relies on the equivalent 
electrical circuit of Fig. \ref{fig01}(b) is described, and the dynamic 
equation for the flux through the ring of the SQUID is derived and normalized.
In Section 3, the dynamic equations for a one-dimensional (1D) SQUID metamaterial
with non-local coupling are derived, and subsequently they are reduced to the 
local coupling limit. In Section 4, various types of chimera states are 
presented and characterized using appropriate measures. In Section 5, the 
possibility to generate and control chimera states with a dc flux gradient, is
explored. A brief discussion and conclusions are presented in Section 6. 

\section{The SQUID oscillator}
The simplest version of a SQUID consists of a superconducting ring interrupted 
by a JJ (Fig. \ref{fig01}(a)), which can be modeled by the equivalent electrical 
circuit of Fig. \ref{fig01}(b); according to that model, the SQUID features 
a self-inductance $L$, a capacitance $C$, a resistance $R$, and a critical 
current $I_c$ which characterizes an ideal JJ. A ``real'' JJ (brown-dashed 
square in Fig. \ref{fig01}(b)) is however modeled as a parallel combination of 
an ideal JJ, the resistance $R$, and the capacitance $C$. When a time-dependent 
magnetic field is applied to the SQUID in a direction transverse to its ring, 
the flux threading the SQUID ring induces two types of currents; the supercurrent, 
which is lossless, and the so-called quasiparticle current which is subject to 
Ohmic losses. The latter roughly corresponds to the current through the branch 
containing the resistor $R$ in Fig. \ref{fig01}(b). The (generally 
time-dependent) flux threading the ring of the SQUID is described in the model
as a flux source, $\Phi_{ext}$. Many variants of SQUIDs have been studied for 
several decades (since 1964) and they have found numerous applications in 
magnetic field sensors, biomagnetism, non-destructive evaluation, and 
gradiometers, among others \cite{Clarke2004a,Clarke2004b}. SQUIDs exhibit very 
rich dynamics including multistability, complex bifurcation structure, and 
chaotic behavior \cite{Hizanidis2018}. 

The magnetic flux $\Phi$ threading the ring of the SQUID is given by
\begin{eqnarray}
\label{eq01}
  \Phi =\Phi_{ext} +L\, I ,
\end{eqnarray}
where $\Phi_{ext}$ is the external flux applied to the SQUID, and
\begin{eqnarray}
\label{eq02}
  I =-C\frac{d^2\Phi}{dt^2} -\frac{1}{R} \frac{d\Phi}{dt} 
     -I_c\, \sin\left(2\pi\frac{\Phi}{\Phi_0}\right), 
\end{eqnarray}
is the total current induced in the SQUID as provided by the resistively and 
capacitively shunted junction (RCSJ) model of the JJ \cite{Likharev1986} (the 
part of the circuit in Fig. \ref{fig01}(b) contained in the brown-dashed square),
$\Phi_0$ is the flux quantum, and $t$ is the temporal variable. The three terms 
in the right-hand-side of Eq. (\ref{eq01}) correspond to the current through the 
capacitor $C$, the current through the resistor $R$, and the supercurrent 
through the ideal JJ, respectively. The combination of Eqs. (\ref{eq01}) and 
(\ref{eq02}) gives 
\begin{eqnarray}
\label{eq03}
  C\frac{d^2\Phi}{dt^2} +\frac{1}{R} \frac{d\Phi}{dt} 
     +I_c\, \sin\left(2\pi\frac{\Phi}{\Phi_0}\right) +\frac{\Phi-\Phi_{ext}}{L}=0. 
\end{eqnarray}
Note that losses decrease with increasing Ohmic resistance $R$, which is a 
peculiarity of the SQUID device. The external flux usually consists of a constant 
(dc) term $\Phi_{dc}$ and a sinusoidal (ac) term of amplitude $\Phi_{ac}$ and 
frequency $\Omega$, i.~e., it is of the form 
\begin{eqnarray}
\label{eq04}
  \Phi_{ext} =\Phi_{dc} +\Phi_{ac} \, \cos( \omega t ).
\end{eqnarray}
The normalized form of Eq. (\ref{eq03}) be obtained by using the relations
\begin{eqnarray}
\label{eq05}
  \phi=\frac{\Phi}{\Phi_0}, ~~~\phi_{ac,dc}=\frac{\Phi_{ac,dc}}{\Phi_0},
  ~~~\tau=\frac{t}{\omega_{LC}^{-1}}, ~~~\Omega=\frac{\omega}{\omega_{LC}},
\end{eqnarray}
where $\omega_{LC} =1 / \sqrt{L C}$ is the inductive-capacitive ($L\, C$) SQUID 
frequency (geometrical frequency), and the definitions
\begin{eqnarray}
\label{eq06}
   \beta=\frac{I_c L}{\Phi_0} =\frac{\beta_L}{2\pi}, \qquad
   \gamma=\frac{1}{R} \sqrt{ \frac{L}{C} }.
\end{eqnarray}
for the rescaled SQUID parameter and the loss coefficient, respectively. Thus we 
get
\begin{eqnarray}
\label{eq07}
   \ddot{\phi} +\gamma \dot{\phi} +\phi +\beta \sin\left( 2\pi \phi \right) =
   \phi_{dc}+ \phi_{ac} \cos(\Omega \tau).
\end{eqnarray}
\begin{figure}[h!]
 \includegraphics[width=8.5cm]{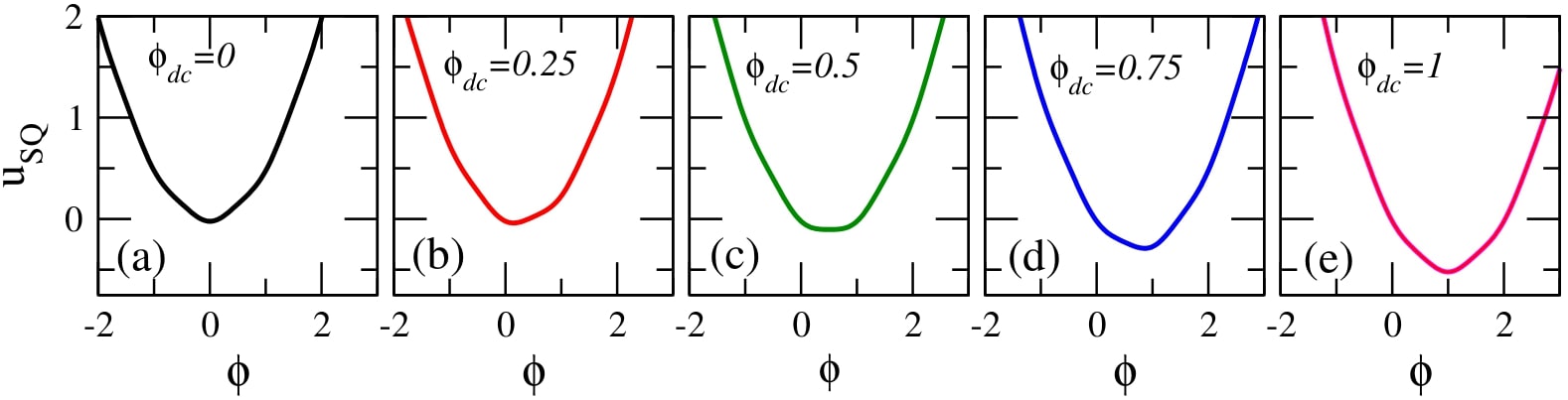} 
\caption{SQUID potential curves $u_{SQ} (\phi)$ for $\beta_L =0.86$, 
 $\phi_{ac} =0$, and (a) $\phi_{dc} =0$; (b) $\phi_{dc} =0.25$; 
 (c) $\phi_{dc} =0.5$; (d) $\phi_{dc} =0.75$; (e) $\phi_{dc} =1.0$.
}
\label{fig02}
\end{figure}
By substituting $\gamma =0$ and $\phi_{ext} =0$ and
$\beta \sin\left( 2\pi \phi \right) \simeq \beta_L \phi$ into Eq. (\ref{eq07}),
we get $\ddot{\phi} +\Omega_{SQ}^2 \phi =0$, with $\Omega_{SQ}=\sqrt{1 +\beta_L}$
being the linear eigenfrequency (resonance frequency) of the SQUID. 
Eq. (\ref{eq07}) can be also written as
\begin{eqnarray}
\label{eq08}
   \ddot{\phi} +\gamma \dot{\phi} =-\frac{d u_{SQ}}{d \phi},
\end{eqnarray}
where
\begin{eqnarray}
\label{eq09}
   u_{SQ} =-\phi_{ext} (\tau) \phi 
    +\frac{1}{2} \left[ \phi^2 -\frac{\beta}{\pi} \cos(2\pi \phi) \right],
\end{eqnarray}
is the normalized SQUID potential, and 
\begin{eqnarray}
\label{eq10}
   \phi_{ext} (\tau) =\phi_{dc} +\phi_{ac} \cos( \Omega \tau ),
\end{eqnarray}
is the normalized external flux. The SQUID potential $u_{SQ}$ given by Eq. 
(\ref{eq09}) is time-dependent for $\phi_{ac} \neq 0$ and $\Omega \neq 0$. 
Here, parameter values of $\beta_L$ less than unity ($\beta_L < 1$) are 
considered, in accordance with recent experiments; in that case, $u_{SQ}$ is a 
single-well, although nonlinear potential. For $\phi_{ext} =\phi_{dc}$, there is 
no time-dependence; however, the shape of $u_{SQ}$ varies with varying 
$\phi_{dc}$, as it can be seen in Fig. \ref{fig02}. The potential $u_{SQ}$ is 
symmetric around a particular $\phi$ for integer and half-integer values of
$\phi_{dc}$. (In Figs. \ref{fig02}(a), (c), and (e), the potential $u_{SQ}$ is
symmetric around $\phi =0$, $0.5$, and $1$, respectively.) For all the other 
values of $\phi_{dc}$, the potential $u_{SQ}$ is asymmetric; this asymmetry of 
$u_{SQ}$ allows for chaotic behavior to appear in an ac and dc driven single 
SQUID through period-doubling bifurcation cascades. Such cascades and the 
subsequent transition to chaos are prevented by a symmetric $u_{SQ}$ which 
renders the SQUID a symmetric system in which period-doubling bifurcations 
are suppressed \cite{Swift1984}. Actually, suppression of period-doubling 
bifurcation cascades due to symmetry occurs in a large class of systems, 
including the sinusoidally driven-damped pendulum.
\begin{figure}[h!]
 \includegraphics[width=8.5cm]{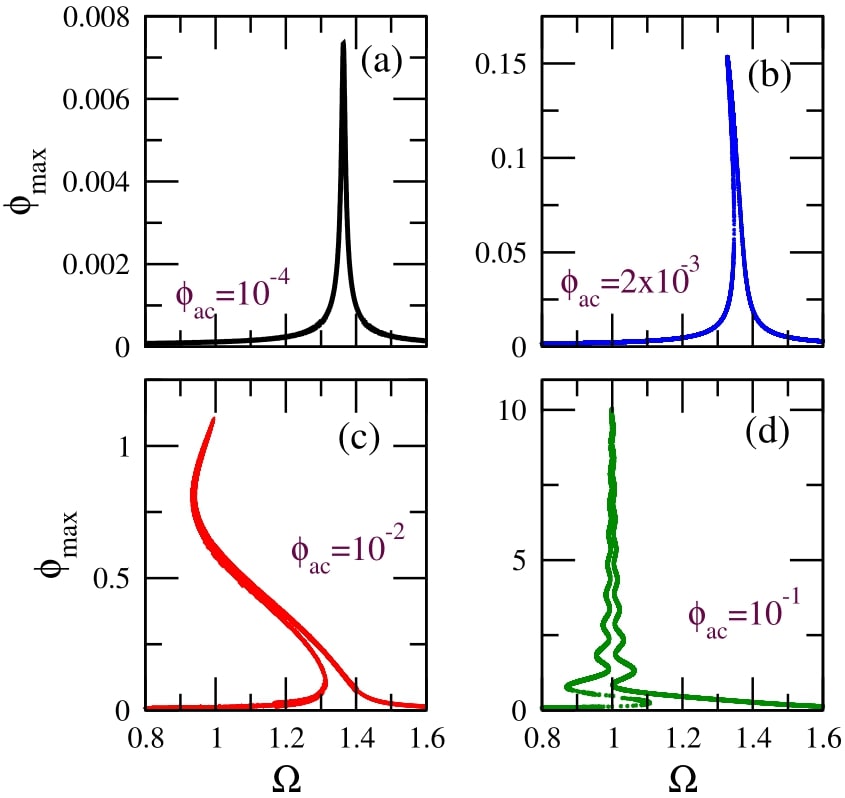} 
\caption{
 Flux amplitude - driving frequency ($\phi_{max} - \Omega$) curves for a SQUID 
 with $\beta_L =0.86$, $\gamma=0.01$, $\phi_{dc} =0$, and 
 (a) $\phi_{ac} =10^{-4}$, (b) $\phi_{ac} =2\times 10^{-3}$, 
 (c) $\phi_{ac} =10^{-2}$, 
 (d) $\phi_{ac} =10^{-1}$.
}
\label{fig03}
\end{figure}

For zero dc flux, the strength of the SQUID nonlinearity increases with increasing ac flux 
amplitude $\phi_{ac}$. This effect is illustrated in Fig. \ref{fig03} in which 
the flux amplitude - driving frequency ($\phi_{max} - \Omega$) curves, i.~e., the
resonance curves, for four values of $\phi_{ac}$ spanning four orders of 
magnitude are shown (for $\phi_{dc}=0$). In Fig. \ref{fig03}(a), for 
$\phi_{ac} =0.0001$, the SQUID is in the linear regime and thus its 
$\phi_{max} - \Omega$ curve is apparently symmetric around the linear SQUID 
eigenfrequency, $\Omega_{SQ} =\sqrt{1 +\beta_L} \simeq 1.364$. Weak nonlinear 
effects begin to appear in Fig. \ref{fig03}(b), for $\phi_{ac} =0.002$, in which 
the curve is slightly bended to the left. In Fig. \ref{fig03}(c), for 
$\phi_{ac} =0.01$, the nonlinear effects are already strong enough to generate a 
multistable $\phi_{max} - \Omega$ curve. In Fig. \ref{fig03}(d), for 
$\phi_{ac} =0.1$, the SQUID is in the strongly nonlinear regime and the 
$\phi_{max} - \Omega$ curve has acquired a {\em snake-like form}. Indeed, the 
curve ``snakes'' back and forth within a narrow frequency region via successive 
saddle-node bifurcations \cite{Hizanidis2018}. Note that in Figs. \ref{fig03}(c) 
and (d), the frequency region with the highest multistability is located around 
the geometrical frequency of the SQUID, i.~e., at $\Omega \simeq 1$ (the $L\, C$ 
frequency in normalized units). Inasmuch the frequency at which $\phi_{max}$ is 
highest can be identified with the ``resonance'' frequency of the SQUID, it can 
be observed that this resonance frequency lowers with increasing $\phi_{ac}$ from 
the linear SQUID eigenfrequency $\Omega_{SQ}$ to the inductive-capacitive 
(geometrical) frequency $\Omega \simeq 1$. Thus, the resonance frequency of the 
SQUID, where its multistability is highest, can be actually tuned by nonlinearity,
i.~e., by varying the ac flux amplitude $\phi_{ac}$. Note that the multistability 
of the SQUID is a purely dynamic effect, which is not related to any local 
minima of the SQUID potential (which is actually single-welled for the values 
of $\beta_L$ considered here, i.~e., for $\beta_L < 1$).    

\begin{figure}[h!]
 \includegraphics[width=8.5cm]{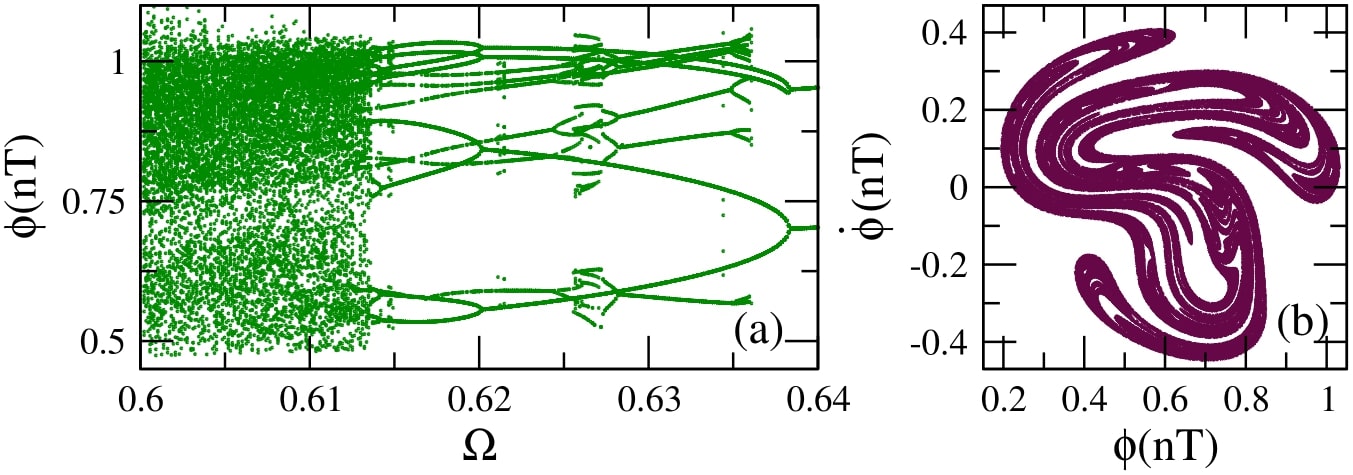} 
\caption{
 (a) Bifurcation diagram of $\phi(nT)$ as a function of the driving frequency 
     $\Omega$, for $\beta_L =0.86$, $\gamma=0.01$, $\phi_{dc} =0.36$, and 
     $\phi_{ac} =0.18$. 
 (b) A typical chaotic attractor on the $\phi - \dot{\phi}$ phase-plane for 
     $\Omega =0.6$. The other parameters are as in (a).
}
\label{fig04}
\end{figure}
For $\phi_{dc} \neq 0$, chaotic behavior appears in wide frequency intervals 
below the geometrical frequency ($\Omega =1$) for relatively high $\phi_{ac}$. 
As it was mentioned above, the SQUID potential $u_{SQ}$ is asymmetric for 
$\phi_{dc} \neq 0$, and thus the SQUID can make transitions to chaos through 
period-doubling cascades \cite{Hizanidis2018}. In the bifurcation diagram shown
in Fig. \ref{fig04}(a), the flux $\phi$ is plotted at the end of each driving 
period $T=2\pi / \Omega$ for several tenths of driving periods (transients have 
been rejected) as a function of the driving frequency $\Omega$. This bifurcation 
diagram reveals multistability as well as a reverse period-doubling cascade 
leading to chaos. That reverse cascade, specifically, begins at $\Omega =0.64$ 
with a stable period-2 solution (i.~e., whose period is two times that of the 
driving period $T$). A period-doubling occurs at $\Omega =0.638$ resulting in a 
stable period-4 solution. The next period-doubling, at $\Omega =0.62$, results 
in a stable period-8 solution. The last period-doubling bifurcation which is 
visible in this scale occurs at $\Omega =0.614$ and results in a stable period-16 
solution. More and more period-doubling bifurcations very close to each other 
lead eventually to chaos at $\Omega =0.6132$. Note that another stable 
multiperiodic solution is present in the frequency interval shown in 
Fig. \ref{fig04}(a). A typical chaotic attractor of the SQUID is shown on the 
$\phi - \dot{\phi}$ phase plane in Fig. \ref{fig04}(b) for $\Omega =0.6$.

\section{SQUID metamaterials: Modelling}
\subsection{Flux dynamics equations}
Consider a one-dimensional periodic arrangement of $N$ identical SQUIDs in a 
transverse magnetic field $\bf H$ as in Fig. \ref{fig01}(c), which center-to-center
distance is $d$ and they are coupled through (non-local) magnetic dipole-dipole 
forces \cite{Lazarides2015b}. The magnetic flux $\Phi_n$ threading the ring of 
the $n-$th SQUID is
\begin{equation}
\label{eq11}
  \Phi_n =\Phi_{ext} +L\, I_n +L\, \sum_{m\neq n} \lambda_{|m-n|} I_m ,
\end{equation}
where $n, m =1,...,N$, $\Phi_{ext}$ is the external flux in each SQUID,
$\lambda_{|m-n|} =M_{|m-n|} / L$ is the dimensionless coupling coefficient between 
the SQUIDs at the sites $m$ and $n$, with $M_{|m-n|}$ being their mutual 
inductance, and  
\begin{eqnarray}
\label{eq12}
    -I_n =C\frac{d^2\Phi_n}{dt^2} +\frac{1}{R} \frac{d\Phi_n}{dt} 
         +I_c\, \sin\left(2\pi\frac{\Phi_n}{\Phi_0}\right) 
\end{eqnarray}
is the current in the $n-$th SQUID as given by the RCSJ model \cite{Likharev1986}. 
The combination of Eqs. (\ref{eq11}) and (\ref{eq12}) gives
\begin{eqnarray}
\label{eq13}
   C\frac{d^2\Phi_n}{dt^2} 
  +\frac{1}{R} \frac{d\Phi_n}{dt}
  +I_c\, \sin\left(2\pi\frac{\Phi_n}{\Phi_0}\right)
\nonumber \\
  +\frac{1}{L} \sum_{m=1}^N  \left( {\bf \hat{\Lambda}}^{-1} \right)_{nm} 
         \left( \Phi_m -\Phi_{ext} \right) =0, 
\end{eqnarray}
where ${\bf \hat{\Lambda}}^{-1}$ is the inverse of the symmetric $N\times N$ 
coupling matrix with elements
\begin{eqnarray}
\label{eq14}
  {\bf \hat{\Lambda}}_{n m} = \left\{ \begin{array}{ll}
        1, & \mbox{if $m= n$};\\
        \lambda_{|m-n|} =\lambda_1 \, |m-n|^{-3}, & \mbox{if $m\neq n$},\end{array} \right.   
\end{eqnarray}
with $\lambda_1$ being the coupling coefficient betwen nearest neighboring SQUIDs.
In normalized form Eq. (\ref{eq13}) reads ($n=1,...,N$)
\begin{eqnarray}
\label{eq15}
  \ddot{\phi}_n +\gamma \dot{\phi}_n +\beta \sin\left( 2\pi \phi_n \right) 
    =\sum_{m=1}^N \left( {\bf \hat{\Lambda}}^{-1} \right)_{nm} 
         \left( \phi_{ext} -\phi_m \right),
\end{eqnarray}
where Eq. (\ref{eq05}) and the definitions Eq. (\ref{eq06}) have been used.
When nearest-neighbor coupling is only taken into account, Eq. (\ref{eq15})
reduces to the simpler form 
\begin{eqnarray}
\label{eq16}
   \ddot{\phi}_n +\gamma \dot{\phi}_n +\phi_n 
   +\beta \sin\left( 2\pi \phi_n \right) =\lambda ( \phi_{n-1} +\phi_{n+1} )
\nonumber \\
+(1 -2\lambda) \phi_{ext},
\end{eqnarray}
where $\lambda=\lambda_1$.

\subsection{Local and nonlocal linear frequency dispersion}
Equation (\ref{eq11}) with $\Phi_{ext} =0$ can be written in matrix form as
\begin{eqnarray}
\label{eq17}
   L\, {\bf \hat{\Lambda}} \vec{I} =\vec{\Phi},
\end{eqnarray}
where the elements of the coupling matrix ${\bf \hat{\Lambda}}$ are given in 
Eq. (\ref{eq14}), and $\vec{I}$, $\vec{\Phi}$ are $N-$dimensional vectors with
components $I_n$, $\Phi_n$, respectively. The linearized equation for the 
current in the $n-$th SQUID, in the lossless case ($R \rightarrow \infty$),
is given from Eq. (\ref{eq12}) as 
\begin{eqnarray}
\label{eq18}
    -\vec{I} = C \frac{d^2}{dt^2} \vec{\Phi} 
               +2\pi \frac{I_c}{\Phi_0} \vec{\Phi},
\end{eqnarray}
where the approximation $\sin(x) \simeq x$ has been employed. By substituting 
Eq. (\ref{eq18}) into Eq. (\ref{eq17}), we get  
\begin{eqnarray}
\label{eq19}
  {\bf \hat{\Lambda}} \left( 
       \frac{1}{\omega_{LC}^2} \frac{d^2}{dt^2} \vec{\Phi} +\beta_L \vec{\Phi}
       \right) +\vec{\Phi} =0.
\end{eqnarray}
In component form, the corresponding equation reads
\begin{eqnarray}
\label{eq20}
  \sum_m {\bf \hat{\Lambda}}_{nm} \left( 
       \frac{1}{\omega_{LC}^2} \frac{d^2}{dt^2} \Phi_m +\beta_L \Phi_m
       \right) +\Phi_n =0,
\end{eqnarray}
or, in normalized form
\begin{eqnarray}
\label{eq21}
  \sum_m {\bf \hat{\Lambda}}_{nm} \left( 
       \frac{1}{\omega_{LC}^2} \ddot{\phi}_m +\beta_L \phi_m
       \right) +\phi_n =0,
\end{eqnarray}
where the overdots denote derivation with respect to the normalized time $\tau$.

Substitute the trial (plane wave) solution
\begin{equation}
\label{eq22}
   \phi_n =\exp^{i (\kappa n -\Omega \tau)},
\end{equation}
where $\kappa$ is the dimensionless wavenumber (in units of $d^{-1}$), into 
Eq. (\ref{eq21}) to obtain
\begin{equation}
\label{eq23}
   \Omega^2 =\frac{1}{S} \left( 1 +\beta_L S \right),  
\end{equation}
where 
\begin{equation}
\label{eq24}
   S =\sum_m {\bf \hat{\Lambda}}_{nm} \exp^{i \kappa (m-n)}.  
\end{equation}
It can be shown that, for the infinite system, the funcion $S$ is
\begin{equation}
\label{eq25}
   S =1 +2 \lambda \sum_{s=1}^\infty \frac{\cos(\kappa s)}{|s|^3}
     =1 +2 \lambda Ci_3 (\kappa),  
\end{equation}
where $s=m-n$, and $Ci_3 (\kappa)$ is a Clausen function. Putting Eq. (\ref{eq25})
into Eq. (\ref{eq23}), we obtain the {\em nonlocal frequency dispersion} for the 1D SQUID 
metamaterial as 
\begin{equation}
\label{eq26}
   \Omega_\kappa =\sqrt{ \frac{\Omega_{SQ}^2 +2\lambda \beta_L Ci_3 (\kappa)}
                              {1 +2\lambda Ci_3 (\kappa)} },
\end{equation}
where $\Omega_{SQ}^2 =1 +\beta_L$. In the case of local (nearest-neighbor) 
coupling the Clausen function $Ci_3 (\kappa)$ is replaced by $\cos(\kappa)$. 
Then, by neglecting terms of order $\lambda^2$ or higher, the {\em local 
frequency dispersion}
\begin{equation}
\label{eq27}
   \Omega_\kappa \simeq \sqrt{ \Omega_{SQ}^2 -2 \lambda \cos(\kappa) }
\end{equation}
is obtained.

\begin{figure}[h!]
\begin{center}
 \includegraphics[width=8.5cm]{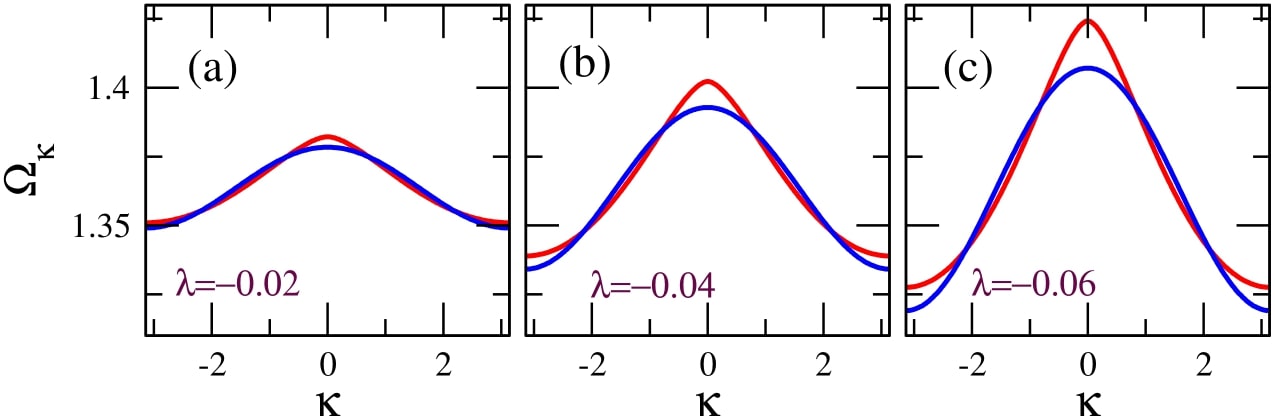} 
\end{center}
\caption{
 Linear frequency dispersion $\Omega =\Omega_\kappa$ for nonlocal (red) and 
 local (blue) coupling, for $\beta_L =0.86$, and (a) $\lambda =-0.02$, 
 (b) $\lambda =-0.04$; (c) $\lambda =-0.06$.  
}
\label{fig06}
\end{figure}
The linear frequency dispersion $\Omega =\Omega_\kappa$, calculated for nonlocal 
and local coupling from Eq. (\ref{eq26}) and (\ref{eq27}), respectively, is 
plotted in Fig. \ref{fig06} for three values of the coupling coefficient $\lambda$. 
The differences between the nonlocal and local dispersion are rather small, 
especially for low  values of $\lambda$, i.~e., for $\lambda =-0.02$ 
(Fig. \ref{fig06}(a)), which are mostly considered here. Although the linear 
frequency bands are narrow, the bandwidth 
$\Delta \Omega =\Omega_{max} -\Omega_{min}$ increases with increasing $\lambda$.
For simplicity, the bandwidth $\Delta \Omega$ can be estimated from Eq. 
(\ref{eq27}); from that equation the minimum and maximum frequencies of the band 
can be approximated by
$ \Omega_{min,max} \simeq \Omega_{SQ} \left( 1 \pm \frac{\lambda}{\Omega_{SQ}^2} \right)$,
so that 
\begin{equation}
\label{eq28}
   \Delta \Omega \simeq \frac{2 |\lambda|}{\Omega_{SQ}}.
\end{equation}
That is, the bandwidth is roughly proportional to the magnitude of $\lambda$.
Note that for physically relevant parameters, the minimum frequency of the
linear band is well above the geometrical (i.~e., inductive-capacitive) frequency
of the SQUIDs in the metamaterial. Thus, for strong nonlinearity, for which the
resonance frequency of the SQUIDs is close to the geometrical one ($\Omega =1$), 
no plane waves can be excited. It is this frequency region where localized and 
other spatially inhomogeneous states such as chimera states are expected to 
emerge (given also the extreme multistability of individual SQUIDs there).

\begin{figure*}[t!]
\begin{center}
 \includegraphics[width=16cm]{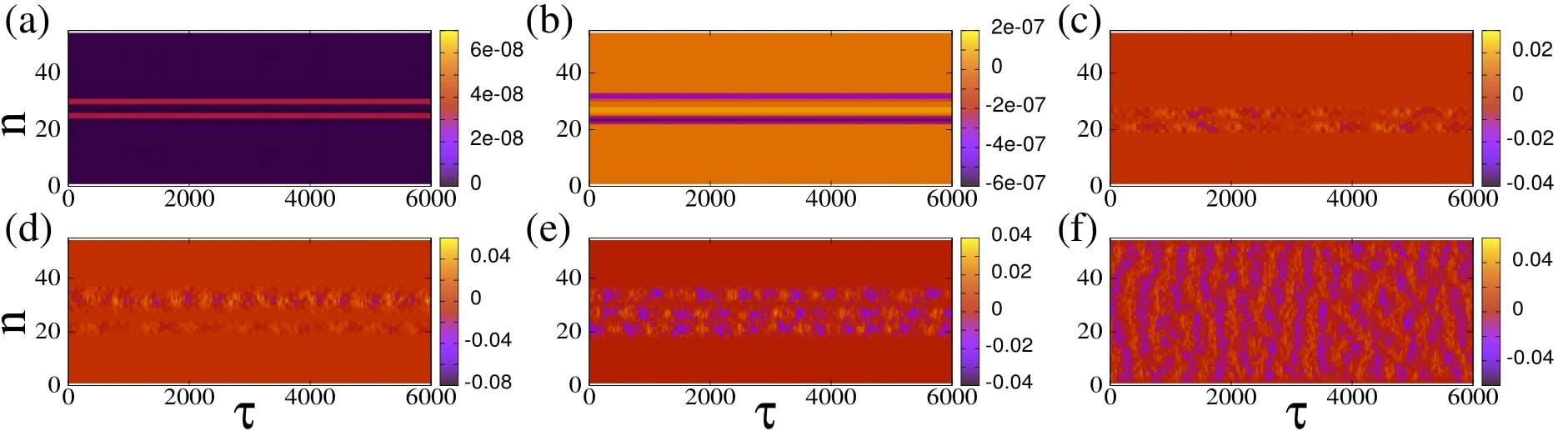} 
\end{center}
\caption{
 Maps of $\langle \dot{\phi}_n \rangle (\tau)$ on the $n - \tau$ plane for 
 $\beta_L =0.86$, $\gamma=0.01$, $\lambda =-0.02$, $\Omega =1.01$, $N=54$,
 $\phi_{dc} =0$, and 
 (a) $\phi_{ac} =0.02$, (b) $\phi_{ac} =0.04$, (c) $\phi_{ac} =0.06$, 
 (d) $\phi_{ac} =0.08$, (e) $\phi_{ac} =0.10$, (f) $\phi_{ac} =0.12$.
}
\label{fig07}
\end{figure*}

\section{Chimeras and other spatially inhomogeneous states}
Eqs. (\ref{eq15}) are integrated numerically in time with free-end boundary 
conditions ($\phi_{N+1} =\phi_0 =0$) using a fourth-order Runge-Kutta algorithm
with time-step $h=0.02$. The initial conditions have been chosen so that they 
lead to chimera states. It should be noted that chimera states can be obtained 
from a huge variety of initial conditions. Here we choose 
\begin{eqnarray}
\label{eq29}
  \phi_n (\tau=0) &=& \left\{ \begin{array}{ll}
          1, & \mbox{for $n_\ell < n \leq n_r$}; \\
          0, & \mbox{otherwise},
                     \end{array} \right.      \\    
  \qquad \dot{\phi}_n(\tau=0) &=&0, 
\end{eqnarray}
with $n_\ell =18$ and $n_r =36$. Eqs. (\ref{eq15}) are first integrated in time 
for a relatively long time-interval, $10^7 ~T$ time-units, where $T=2\pi/\Omega$ 
is the driving period, so that the system has reached a steady-state. While the 
SQUID metamaterial is in the steady-state, Eqs. (\ref{eq15}) are integrated for 
$\tau_{sst} =1000 ~T$ more time-units. Then, the profiles of the time-derivatives 
of the fluxes, averaged over the driving period $T$, i.~e.,  
\begin{equation}
\label{eq30}
  {\langle \dot{\phi}_n \rangle}_T =\frac{1}{T} \int_0^T \dot{\phi}_n \, d\tau,
  ~~~~ n=1,...,N,
\end{equation}
are mapped as a function of $\tau$. Such maps are shown in Fig. \ref{fig07}, 
for several values of the ac flux amplitude, $\phi_{ac}$. In these maps, areas 
with uniform colorization indicate that the SQUID oscillators there are 
synchronized, while areas with non-uniform colorization indicate that they are
desynchronized.

\begin{figure}[t!]
\begin{center}
 \includegraphics[width=8cm]{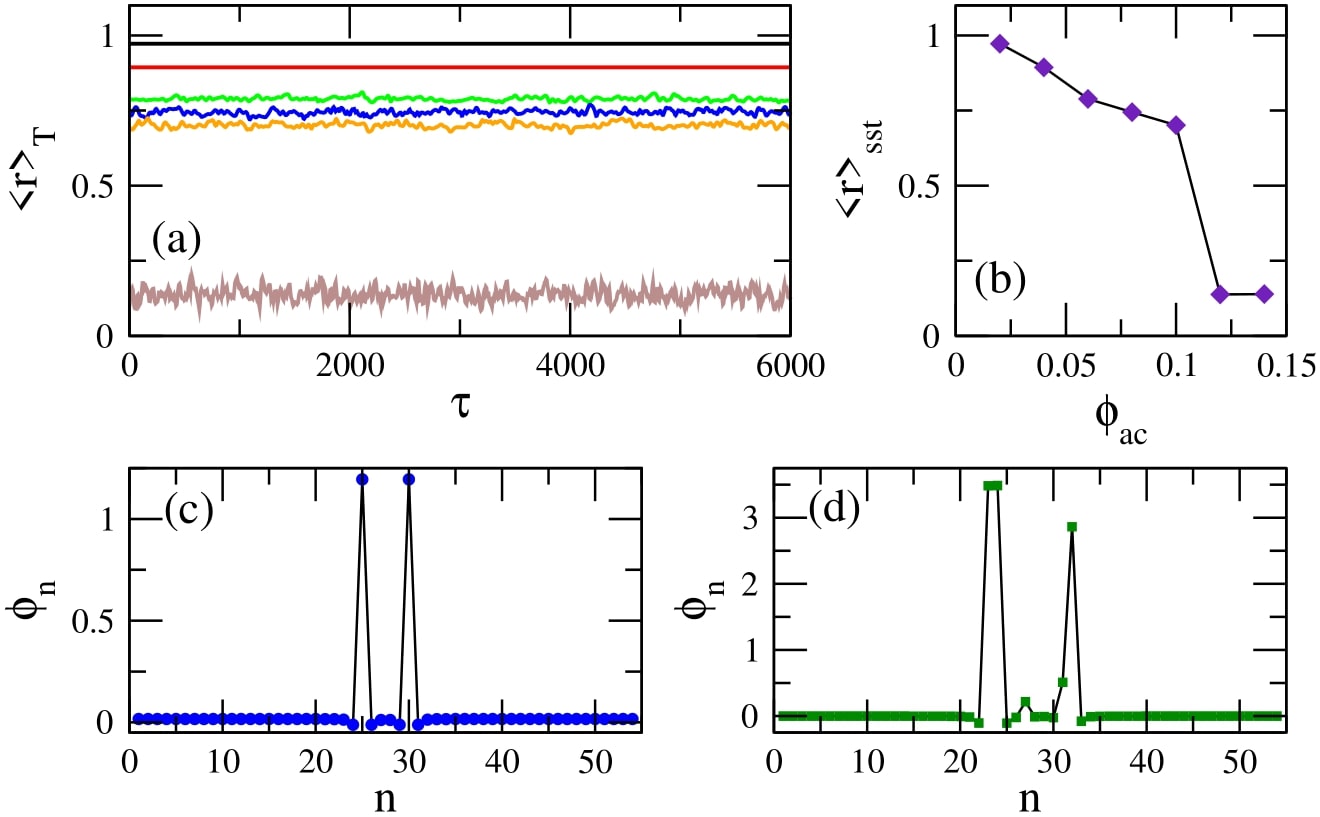} 
\end{center}
\caption{
 (a) The magnitude of the synchronization parameter averaged over the driving 
     period, ${\langle r \rangle}_T$, as a function of time $\tau$ for 
     $\beta_L =0.86$, $\gamma=0.01$, $\lambda =-0.02$, $\Omega =1.01$, $N=54$,
     $\phi_{dc} =0$, and $\phi_{ac} =0.02$ (black), $\phi_{ac} =0.04$ (red), 
     $\phi_{ac} =0.06$ (green), $\phi_{ac} =0.08$ (blue), $\phi_{ac} =0.10$ 
     (orange), $\phi_{ac} =0.12$ (brown).
 (b) The magnitude of the synchronization parameter averaged over the 
     steady-state integration time $\tau_{sst}$, ${\langle r \rangle}_{sst}$, as 
     a function of the ac flux amplitude $\phi_{ac}$. The other parameters are
     as in (a).
 (c) The flux profile $\phi_n$ for $\phi_{ac} =0.02$ and the other parameters
     as in (a).
 (d) The flux profile $\phi_n$ for $\phi_{ac} =0.04$ and the other parameters
     as in (a). 
}
\label{fig08}
\end{figure}

In Figs. \ref{fig07}(a) and (b), i.~e., for low values of $\phi_{ac}$, chimera 
states are not excited since the ${\langle \dot{\phi}_n \rangle}_T$ are practically 
zero during the steady-state integration time. However, this does not mean that 
the state of the SQUID metamaterial is spatially homogeneous, as we shall see 
below. For higher values of $\phi_{ac}$, chimera states begin to appear, in which 
one or more desynchronized clusters of SQUID oscillators roughly in the middle 
of the SQUID metamaterial are visible (Figs. \ref{fig07}(c)-(e)). For even higher 
values of $\phi_{ac}$, as can be seen in Fig. \ref{fig07}(f), the whole SQUID 
metamaterial is desynchronized. In order to quantify the degree of synchronization 
for SQUID metamaterials at a particular time-instant $\tau$, the magnitude of 
the complex synchronization (Kuramoto) parameter $r$ is calculated, where   
\begin{equation}
\label{eq31}
  r(\tau) =\left| \Psi (\tau)  \right| 
          =\frac{1}{N} \left| \sum_n e^{2\pi i \phi_n (\tau)} \right|. 
\end{equation} 
Below, two averages of $r (\tau)$ are used for the characterization of a
particular state of SQUID metamaterials, i.~e., the average of $r (\tau)$ over 
the driving period $T$, ${\langle r \rangle}_T (\tau)$, and the average of 
$r (\tau)$ over the steady-state integration time ${\langle r \rangle}_{sst}$. 
These are defined, respectively, as
\begin{equation}
\label{eq32}
   \left< r (\tau) \right>_T =\frac{1}{T} \int_0^T r (\tau)\, d\tau, \qquad
   {\langle r \rangle}_{sst} =\frac{1}{\tau_{sst}} \int_0^{\tau_{sst}} r (\tau)\, d\tau. 
\end{equation} 
The calculated ${\langle r \rangle}_T (\tau)$ for the states shown in 
Fig. \ref{fig07}, clarify further their nature. In Fig. \ref{fig08}(a), 
${\langle r \rangle}_T (\tau)$ is plotted as a function of time $\tau$ for all the 
six states presented in Fig. \ref{fig07}. It can be seen that for 
$\phi_{ac} =0.02$ and $0.04$ (black and red curves), calculated for the states 
of the SQUID metamaterial in Figs. \ref{fig07}(a) and (b), respectively, 
${\langle r \rangle}_T (\tau)$ is constant in time, although less than unity. 
For such states, ${\langle r \rangle}_T (\tau) ={\langle r \rangle}_{sst}$, where 
${\langle r \rangle}_{sst}$ can be inferred from Fig. \ref{fig08}(b) for the curves 
of interest to be ${\langle r \rangle}_{sst} \simeq 0.972$ and 
${\langle r \rangle}_{sst} \simeq 0.894$ for $\phi_{ac} =0.02$ and $0.04$, 
respectively. The lack of fluctuations indicates that these states consist of 
``clusters'' in which the SQUID oscillators are synchronized together. However, 
the clusters are not synchronized to each other, resulting in a partially 
synchronized state with ${\langle r \rangle}_T (\tau) < 1$. The exact nature of 
these partially synchronized states can be clarified by plotting the flux 
profiles $\phi_n$ at the end of the steady-state integration time as shown in 
Figs. \ref{fig08}(c) and (d). In these figures, it can be observed that all but 
a few SQUID oscillators are synchronized; in addition, those few SQUIDs execute 
high-amplitude flux oscillations. Moreover, it has been verified that the 
frequency of all the flux oscillations is that of the driving, $\Omega$. Such 
states can be classified as discrete breathers/multi-breathers, i.~e., spatially 
localized and time-periodic excitations which have been proved to emerge 
generically in nonlinear networks of weakly coupled oscillators \cite{Flach2008a}. 
In the present case, the multibreathers shown in Figs. \ref{fig08}(c) and (d) 
can be further characterized as {\em dissipative} ones \cite{Flach2008b}, since 
they emerge through a delicate balance of input power and intrinsic losses. They 
have been investigated in some detail in SQUID metamaterials in one and two 
dimensions \cite{Lazarides2008a,Tsironis2009,Lazarides2012,Lazarides2015a},
as well as in SQUID metamaterials on two-dimensional Lieb lattices 
\cite{Lazarides2018a}. 

The corresponding ${\langle r \rangle}_T (\tau)$ for the states shown in Figs. 
\ref{fig07}(c), (d), (e), and (f), are shown in \ref{fig08}(a) as green, blue, 
orange, and brown curves, respectively. In these curves there are apparently 
fluctuations around their temporal average over the steady-state integration 
time (shown in Fig. \ref{fig08}(b)). These fluctuations are typically associated 
with the level of metastability of the chimera states 
\cite{Shanahan2010,Wildie2012}; an appropriate measure of metastability for SQUID 
metamaterials is the full-width half-maximum (FWHM) of the distribution of 
${\langle r \rangle}_T$ \cite{Lazarides2015b}. The FWHM can be used to compare the 
metastability levels of different chimera states. For synchronized (spatially 
homogeneous) and partially synchronized states such as those in Figs. 
\ref{fig07}(a) and (b), the FWHM of the corresponding distribution of the values 
of ${\langle r \rangle}_T$ is practically zero.  

\begin{figure*}[t!]
\begin{center}
\includegraphics[width=16cm]{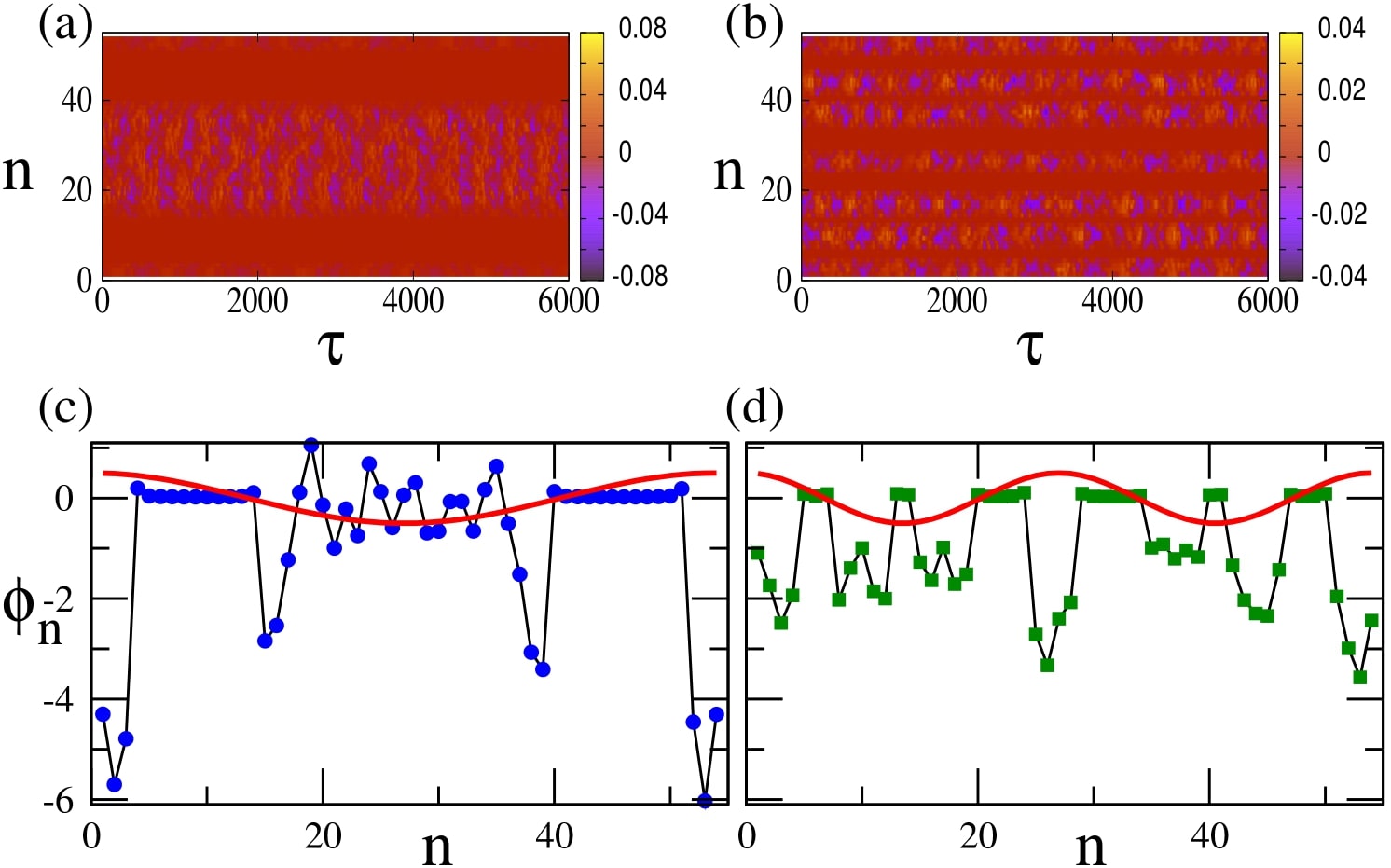} 
\end{center}
\caption{
 (a) Map of ${\langle \dot{\phi}_n \rangle}_T$ on the $n - \tau$ plane for 
 $\beta_L =0.86$, $\gamma=0.01$, $\lambda =-0.02$, $\Omega=1.01$, $N=54$, 
 $\phi_{dc} =0$, $\phi_{ac} =0.1$, and initial conditions given by 
 Eq. (\ref{eq33}) with $j=1$.
 (b) Same as in (a) with initial conditions given by Eq. (\ref{eq33}) with $j=2$.
 (c) Flux profile $\phi_n$ at the end of the steady-state integration time 
     (blue circles, the black line is a guided to the eye), obtained with the 
     initial conditions Eq. (\ref{eq33}) with $j=1$ (red curve). 
 (d) Flux profile $\phi_n$ at the end of the steady-state integration time 
     (blue circles, the black line is a guided to the eye), obtained with the 
     initial conditions Eq. (\ref{eq33}) with $j=2$ (red curve). 
}
\label{fig09}
\end{figure*}

Another set  of initial conditions which gives rise to chimera states is of the 
form \cite{Hizanidis2016a} 
\begin{equation}
\label{eq33}
   \phi_n (\tau =0) =\frac{1}{2} \cos \left( \frac{2 j \pi n}{N} \right), 
   \qquad \dot{\phi}_n (\tau =0) =0,
\end{equation} 
where $n=1,...,N$. 
The initial conditions in Eq. (\ref{eq33}) allow for generating multiclustered 
chimera states, in which the number of clusters depends on $j$. In Figs. 
\ref{fig09}(a) and (b), maps of ${\langle \dot{\phi}_n \rangle}_T$ on the $n-\tau$ 
plane for $j=1$ and $j=2$, respectively, are shown. In Fig. \ref{fig09}(a), 
three large clusters can be distinguished; in the two of them, the SQUID 
oscillators are synchronized, while in the third one, in between the two 
sychronized clusters, the SQUID oscillators are desynchronized. The flux profile 
$\phi_n$ of that state at the end of the steady-state integration time 
$\tau_{sst} =6000$, is shown in Fig. \ref{fig09}(c) as blue circles (the black 
curve is a guide to the eye) along with the initial condition (red curve). 
It can be seen that two more desynchronized clusters at the ends of the 
metamaterial, which are rather small (they consist of only a few SQUIDs each), 
are visible. Obviously, the synchronized clusters correspond to the spatial 
interval indicated by the almost horizontal segments in the $\phi_n$ profile. 
The corresponding ${\langle \dot{\phi}_n \rangle}_T$ map and flux profile $\phi_n$ 
for $j=2$ is shown in Figs. \ref{fig09}(b) and (d), respectively. In this case, 
a number of six (6) synchronized clusters and seven (7) desynchronized clusters 
are visible in both Figs. \ref{fig09}(b) and (d). In Figs. \ref{fig09}(d), the
red curve is the initial condition from Eq. (\ref{eq33}) with $j=2$. Chimera 
states with even more ``heads'' can be generated from the initial condition Eq.
(\ref{eq33}) for $j > 2$ in larger systems (here $N=54$).

\begin{figure*}[t!]
\begin{center}
 \includegraphics[width=16cm]{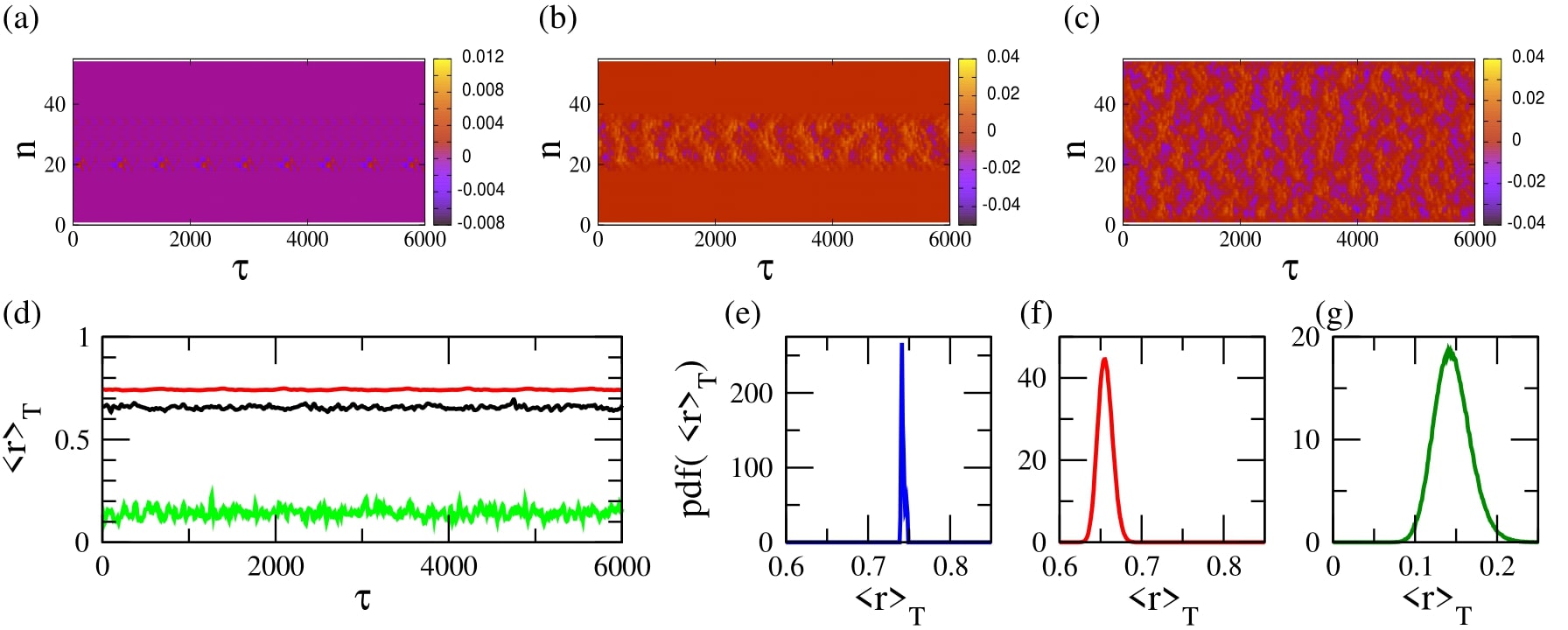} 
\end{center}
\caption{
 (a) Map of ${\langle \dot{\phi}_n \rangle}_T$ on the $n - \tau$ plane for 
     $\beta_L =0.86$, $\gamma=0.01$, $\lambda =-0.02$, $\Omega =1.01$, $N=54$,
     $\phi_{dc} =0$, $\phi_{ac} =0.06$, and initial conditions given by Eq. 
     (\ref{eq29}).
 (b) Same as in (a) but with $\phi_{ac} =0.10$.
 (c) Same as in (a) and (b) but with $\phi_{ac} =0.12$.
 (d) The magnitude of the synchronization parameter averaged over the driving
     period, ${\langle r \rangle}_T$, as a function of time $\tau$ for 
     $\phi_{ac} =0.06$ (red), $\phi_{ac} =0.1$ (black), and $\phi_{ac} =0.12$ 
     (green). The other parameters are as in (a).
 (e) The distribution of $10^6$ values of ${\langle r \rangle}_T$, 
     $pdf( {\langle r \rangle}_T )$, for the chimera state shown in (a).
 (f) Same as in (e) for the chimera state shown in (b).
 (g) Same as in (e) and (f) for the completely desynchronized state shown in (c).
}
\label{fig10}
\end{figure*}

Similar chimera states can be generated with local (nearest-neighbor) coupling 
between the SQUIDs of the metamaterial. For that purpose, Eq. (\ref{eq16}) is 
integrated in time using a fourth order Runge-Kutta algorithm with free-end 
boundary conditions and the initial conditions of Eq. (\ref{eq29}). As above, in 
order to eliminate transients and reach a steady-state, Eq. (\ref{eq16}) is 
integrated for $10^7~T$ time units and the results are discarded. Then, 
(\ref{eq16}) is integrated for $\tau_{sst} =10^3~T$ more time units 
(steady-state integration time), and ${\langle \dot{\phi}_n \rangle}_T$ is mapped 
on the $n-\tau$ plane (Fig. \ref{fig10}). The emerged states are very similar to 
those shown in Fig. \ref{fig07}, which is the case of non-local coupling between 
the SQUIDs. In particular, the states shown in Fig. \ref{fig10}(a), (b), and (c), 
have been generated for exactly the same parameters and initial-boundary 
conditions as those in Fig. \ref{fig07}(c), (e), and (f), respectively, i.e, for 
$\phi_{ac} =0.06$, $0.1$, and $0.12$. Note that the state of the SQUID 
metamaterial for $\phi_{ac} =0.12$ is completely desynchronized both in Fig.
\ref{fig07}(f) and \ref{fig10}(c). One may also compare the plots of the 
corresponding ${\langle r \rangle}_T$ as a function of $\tau$, which are shown in 
Fig. \ref{fig10}(d) for the local coupling case. The averages of $r$ over the 
steady-state integration time $\tau_{sst}$ for $\phi_{ac} =0.06$, $0.1$, $0.12$ 
are, respectively, ${\langle r \rangle}_{sst} =0.757$, $0.656$, $0.136$ for the 
nonlocal coupling case and ${\langle r \rangle}_{sst} =0.743$, $0.656$, $0.146$ for 
the local coupling case. The probability distribution function of the values of 
${\langle r \rangle}_T$, $pdf( {\langle r \rangle}_T )$, for the three states in Figs. 
\ref{fig10}(a)-(c) are shown in Figs. \ref{fig10}(e)-(g), respectively. As it 
was mentioned above, the FWHM of such a distribution is a measure of the 
metastability of the corresponding chimera state. The FWHM for the distributions 
in Fig. \ref{fig10}(e) and (f), calculated for the chimera states shown in Fig. 
\ref{fig10}(a) and (b), are respectively $0.003$ and $0.0215$. Thus, it can be 
concluded that the chimera state of Fig. \ref{fig10}(b) is more metastable than 
that in Fig. \ref{fig10}(a). The distribution in Fig. \ref{fig10}(g) has a FWHM
much larger than the ones of the distributions in Figs. \ref{fig10}(e) and (f)
as expected, since it has been calculated for the completely desynchronized 
state of Fig. \ref{fig10}(c). Note that $10^6$ values of ${\langle r \rangle}_T$ 
have been used to obtain each of the three distributions. Also, these 
distributions are normalized such that their area sums to unity.   

The chimera states do not result from destabilization of the synchronized state 
of the SQUID metamaterial; instead, they coexist with the latter, which can be 
reached simply by integrating the relevant flux dynamics equations with zero 
initial conditions, i.~e., with $\phi_n (\tau =0) =0$ and 
$\dot{\phi}_n (\tau =0) =0$ for any $n$. In order to reach a chimera state, on 
the other hand, appropriately chosen initial conditions such as those in Eq. 
(\ref{eq29}) or Eq. (\ref{eq33}) have to be used.  However, one cannot expect 
that the synchronized state is stable over the whole external parameter space, 
i.~e., the ac flux amplitude $\phi_{ac}$, the frequency of the ac flux field 
$\Omega$, and the dc flux bias $\phi_{dc}$. In order to explore the stability of 
the synchronized state of the SQUID metamaterial, the magnitude of the 
synchronization parameter averaged over the steady-state integration time, 
${\langle r \rangle}_{sst}$, is calculated and then mapped on the 
$\phi_{dc} - \phi_{ac}$ parameter plane. For each pair of $\phi_{ac}$ and 
$\phi_{dc}$ values, the SQUID metamaterial is initialized with zeros (it is at
``rest''). Once again, the frequency $\Omega$ is chosen to be very close to the 
geometrical resonance $\Omega_{LC}$ ($\Omega \simeq 1$). In Fig. \ref{fig11}, 
maps of ${\langle r \rangle}_{sst}$ on the $\phi_{dc} - \phi_{ac}$ plane are shown 
for four driving frequencies $\Omega$ around unity. These maps are a kind of 
{\em ``synchronization phase diagrams''}, in which ${\langle r \rangle}_{sst} =1$ 
indicates a synchronized state while ${\langle r \rangle}_{sst} <1$ indicates a 
partially or completely desynchronized state. In all subfigures, but perhaps 
most clearly seen in Fig. \ref{fig11}(c) (for $\Omega =1.01$) there are abrupt
transitions between completely synchronized (red areas) and completely 
desynchronized (light blue areas) states. It can be verified by inspection of
the flux profiles (not shown) that these {\em synchronization-desynchronization 
transitions} do not go through a stage in which chimera states are generated; 
instead, the destabilization of a synchronized state results either in a 
completely desynchronized state (light blue areas) or a clustered state (green 
areas). Thus, it seems that chimera states cannot be generated when the SQUID
metamaterial is initially at ``rest'', i.~e., with zero initial conditions. As we 
shall see in the next Section (Section 5), this is not true for a 
position-dependent external flux $\phi_{ext} =\phi_{ext} (n)$.  

\section{Chimera generation by dc flux gradients}
\subsection{Modified flux dynamics equations}
\begin{figure}[h!]
\begin{center}
 \includegraphics[width=8.5cm]{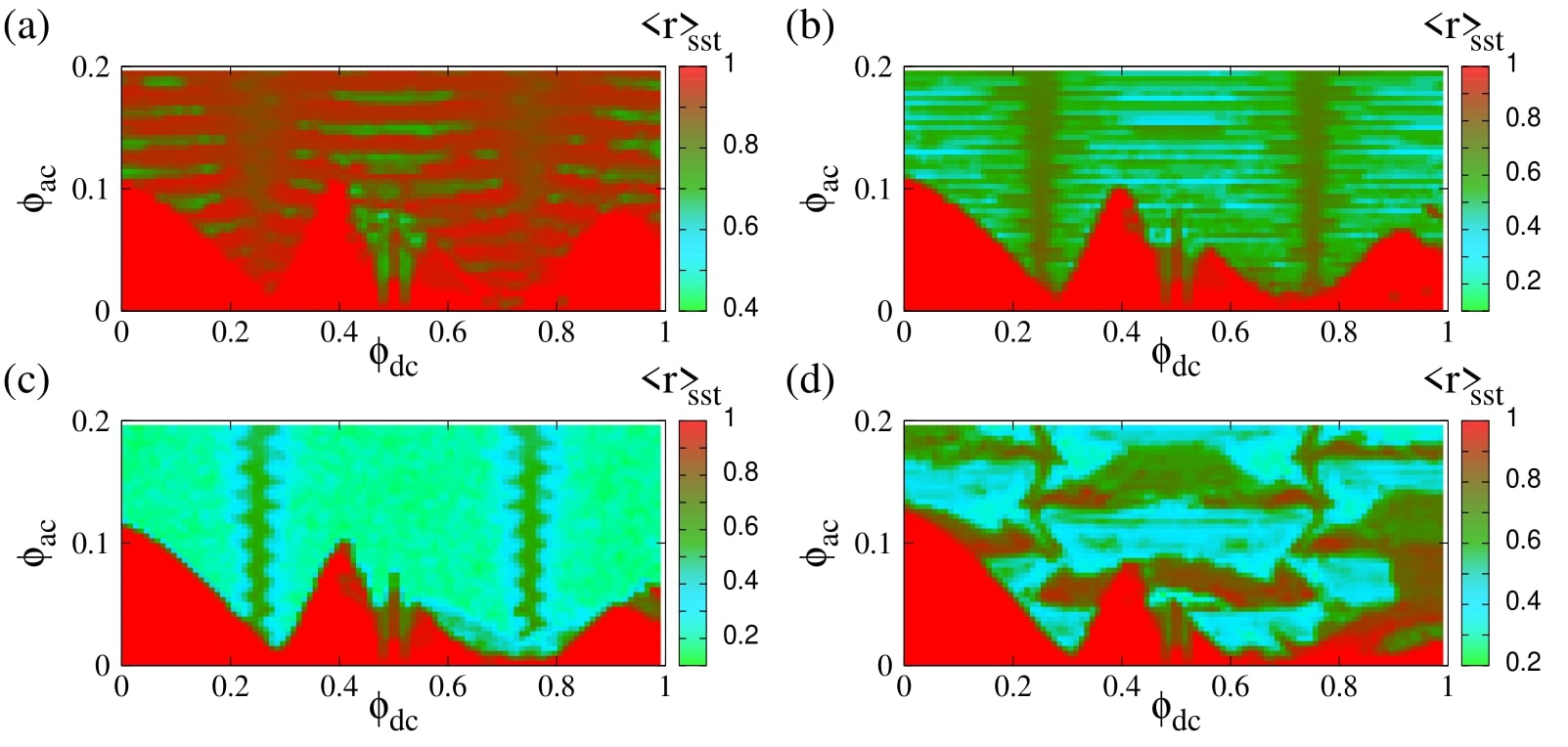} 
\end{center}
\caption{
 Map of the magnitude of the synchronization parameter averaged over the 
 steady-state integration time, ${\langle r \rangle}_{sst}$, on the dc flux bias - 
 ac flux amplitude ($\phi_{dc} -\phi_{ac}$) parameter plane, for $\beta_L =0.86$, 
 $\gamma=0.01$, $\lambda =-0.02$, $N=54$, and 
 (a) $\Omega =1.03$, (b) $\Omega =1.02$, (c) $\Omega =1.01$, (d) $\Omega =0.982$.
}
\label{fig11}
\end{figure}

In obtaining the results of Fig. \ref{fig11}, a spatially homogeneous dc flux 
$\phi_{dc}$ over the whole SQUID metamaterial is considered. Although, all the 
chimera states presented here are generated at $\phi_{dc} =0$, such states can 
be also generated in the presence of a spatially constant, non-zero $\phi_{dc}$,
by using appropriate initial conditions (not shown here). In this Section,
the generation of chimera states in SQUID metamaterials driven by an ac flux and 
biased by a dc flux gradient is demonstrated, for the SQUID metamaterial being
initially at ``rest''. The application of a dc flux gradient along the SQUID 
metamaterial is experimentally feasible with the set-up of Ref. \cite{Zhang2015}. 
Consider the SQUID metamaterial model in Section $3.1$ in the case of local 
coupling (for simplicity), in which the dc flux is assumed to be 
position-dependent, i.~e., $\phi_{dc} =\phi_n^{dc}$. Then, Eqs. (\ref{eq16}) can 
be easily modified to become
\begin{eqnarray}
\label{eq50}
  \ddot{\phi}_n +\gamma \dot{\phi}_n +\phi_n
   +\beta\, \sin( 2 \pi \phi_n ) =\phi_n^{eff} (\tau)
\nonumber \\
   +\lambda ( \phi_{n-1} +\phi_{n+1} ),
\end{eqnarray}	
where
\begin{eqnarray}
\label{eq51}
\phi_n^{eff} =\phi_n^{ext} -\lambda ( \phi_{n-1}^{ext} +\phi_{n+1}^{ext} ),
\end{eqnarray}
with
\begin{eqnarray}
\label{eq52}
  \phi_n^{ext} = \phi_n^{dc}  +\phi_{ac} \cos(\Omega \tau ).
\end{eqnarray}
In the following, the dc flux function $\phi_n^{dc}$ is assumed to be of the
form
\begin{equation}
\label{eq53}
  \phi_n^{dc} =\frac{n-1}{N-1} \phi_{max}^{dc}, ~~~~~~ n=1, ... ,N,
\end{equation}
so that the dc flux bias increases linearly from zero (for the SQUID at $n=1$) 
to $\phi_{max}^{dc}$ (for the SQUID at the $n=N$).

\subsection{Generation and control of chimera states}
Equations (\ref{eq50}) are integrated numerically in time with free-end boundary 
conditions (Eqs. (\ref{eq50})) using a fourth-order Runge-Kutta algorithm with 
time-step $h=0.02$. The SQUID metamaterial is initially at ``rest'', i.~e.,
\begin{equation}
\label{eq54}
  \phi_n (\tau=0) =0, \qquad \dot{\phi}_n (\tau=0) =0, ~~~~ n=1,...,N. 
\end{equation}
This system is integrated for $10^5 ~T$ time units to eliminate the transients
and then for more $\tau_{sst} =10^5 ~T$ time units during which the temporal 
averages ${\langle r \rangle}_{sst}$ and ${\langle r \rangle}_T (\tau)$ are calculated. 
Note that the transients die-out faster in this case since the SQUID 
metamaterial is initialized with zeros.
\begin{figure}[h!]
\begin{center}
 \includegraphics[width=8.5cm]{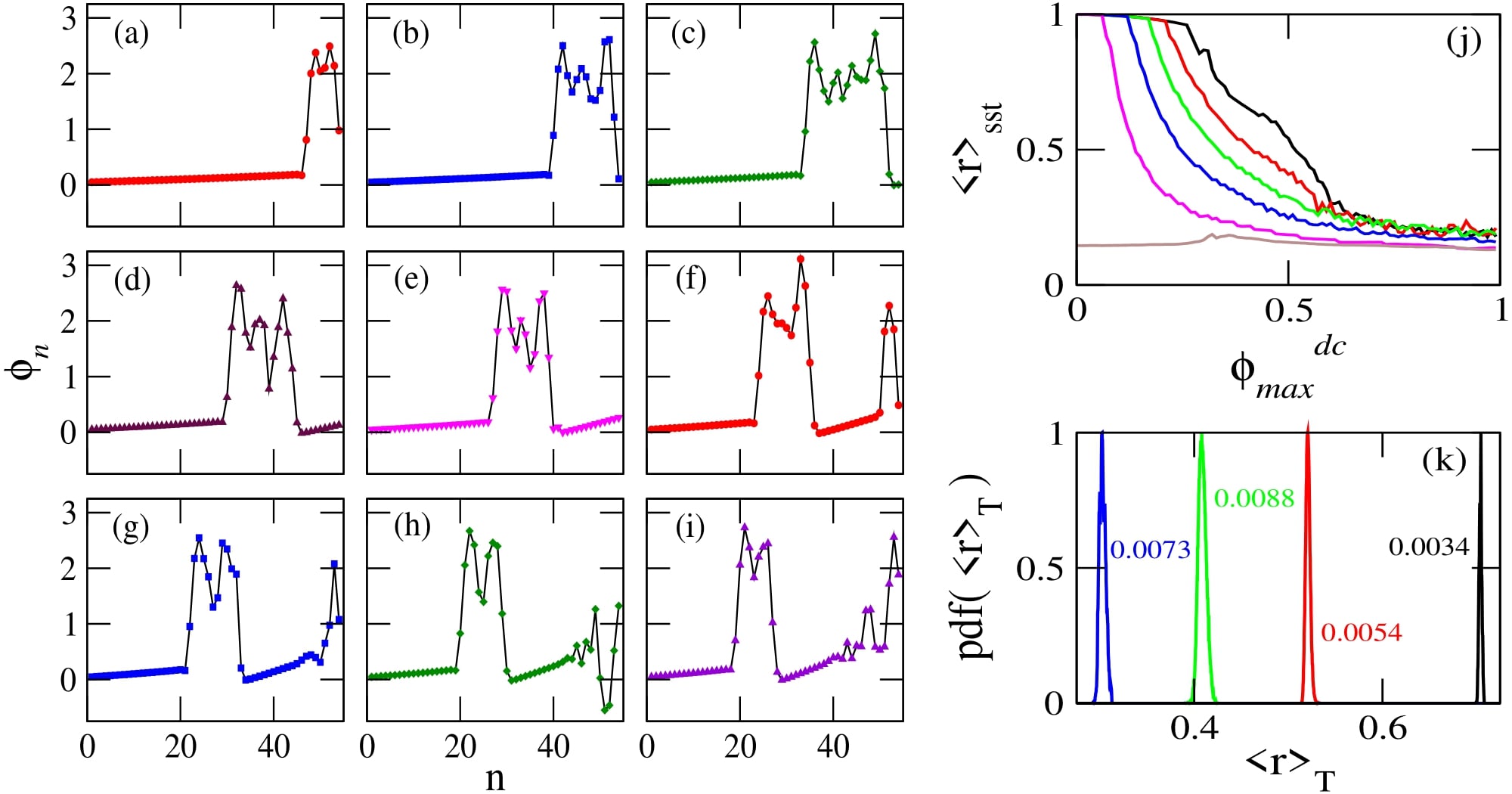} 
\end{center}
\caption{
 Flux profiles $\phi_n$ as a function of $n$ for $\beta_L =0.86$, $\gamma=0.01$, 
 $\lambda =-0.02$, $N=54$, $\phi_{ac} =0.04$, $\Omega =1.01$, and 
 (a) $\phi_{max}^{dc} =0.25$; (b) $0.30$; (c) $0.35$; (d) $0.40$; (e) $0.45$; 
 (f) $0.50$; (g) $0.55$; (h) $0.60$; (i) $0.65$.
 (j) The magnitude of the synchronization parameter averaged over the 
     steady-state integration time ${\langle r \rangle}_{sst}$ as a function of 
     $\phi_{max}^{dc}$ for the parameters of (a)-(i) but with $\phi_{ac} =0.02$ 
     (black), $0.04$ (red), $0.06$ (green), $0.08$ (blue), $0.10$ (magenta), 
     $0.12$ (brown).
 (k) Distributions of the values of ${\langle r \rangle}_T$ for $\phi_{ac} =0.04$, 
     and $\phi_{max}^{dc} =0.30$ (black), $0.40$ (red), $0.50$ (green), $0.60$
     (blue). The other parameters as in (a)-(i). The numbers next to the 
     distributions are the corresponding full-width half-maximums.
}
\label{fig12}
\end{figure}
Typical flux profiles $\phi_n$, plotted at the end of the steady-state 
integration time are shown in Figs. \ref{fig12}(a)-(i). The varying parameter in 
this case is $\phi_{max}^{dc}$, which actually determines the gradient of the dc
flux. The state of the SQUID metamaterial remains {\em almost homogeneous in 
space} for $\phi_{max}^{dc}$ increasing from zero to $\phi_{max}^{dc} =0.22$. At 
that critical value of $\phi_{max}^{dc}$, the spatially homogeneous (almost 
synchronized) state breaks down, for several SQUIDs close to $n=N$ become 
desynchronized with the rest (because the dc flux is higher at this end). The 
number of desynchronized SQUIDs for $\phi_{max}^{dc} =0.25$ is about $6-7$ (Fig. 
\ref{fig12}(a)). For further increasing $\phi_{max}^{dc}$, more and more SQUIDs 
become desynchronized, until they form a well-defined desynchronized cluster 
(Fig. \ref{fig12}(b) for $\phi_{max}^{dc} =0.30$). As $\phi_{max}^{dc}$ continues 
to increase, the desynchronized cluster clearly shifts to the left, i.~e., towards 
$n=1$ (Fig. \ref{fig12}(c)-(e)). Further increase of $\phi_{max}^{dc}$ generates 
a second desynchronized cluster around $n=N$ for $\phi_{max}^{dc} =0.50$ 
(Fig. \ref{fig12}(f)), which persists for values of $\phi_{max}^{dc}$ at least 
up to $0.65$. With the formation of the second desynchronized cluster, the first 
one clearly becomes smaller and smaller with increasing $\phi_{max}^{dc}$ (see 
Figs. \ref{fig12}(f)-(i)). Above, the expression ``almost homogeneous'' was used
instead of simply ``homogeneous'', because complete homogeneity is not possible 
due to the dc flux gradient. However, for $\phi_{max}^{dc} < 0.22$, the degree 
of homogeneity (synchronization) is more than $99 \%$, i.~e., the values of the 
synchronization parameter ${\langle r \rangle}_{sst}$ are higher than $0.99$ 
(${\langle r \rangle}_{sst} > 0.99$). The dependence of ${\langle r \rangle}_{sst}$ on 
$\phi_{max}^{dc}$ for several values of the ac flux amplitude $\phi_{ac}$ is 
shown in Fig. \ref{fig12}(j). The SQUID metamaterial remains in an almost 
synchronized state (with ${\langle r \rangle}_{sst} > 0.96$ below a critical value 
of $\phi_{max}^{dc}$, which depends on the ac flux amplitude $\phi_{ac}$. That 
critical value of $\phi_{max}^{dc}$ is lower for higher $\phi_{ac}$. For values 
of $\phi_{max}^{dc}$ higher than the critical one, ${\langle r \rangle}_{sst}$ 
gradually decreases until it saturates at ${\langle r \rangle}_{sst} \simeq 0.12$. 
For $\phi_{ac} =0.12$, the SQUID metamaterial is in a completely desynchronized 
state for any value of $\phi_{max}^{dc}$ (brown curve). The distributions of the 
values of ${\langle r \rangle} _T$, obtained during the steady-state integration 
time, are shown in Fig. \ref{fig12}(k) for $\phi_{max}^{dc} =0.30$ (black), 
$0.40$ (red), $0.50$ (green), and $0.60$ (blue). As expected, the maximum of the 
distributions shifts to lower ${\langle r \rangle} _T$ with increasing 
$\phi_{max}^{dc}$. These distributions have been divided by their maximum value 
for easiness of presentation, and the number next to each distribution is its 
full-width half-maximum (FWHM).
\begin{figure}[h!]
\begin{center}
 \includegraphics[width=8.5cm]{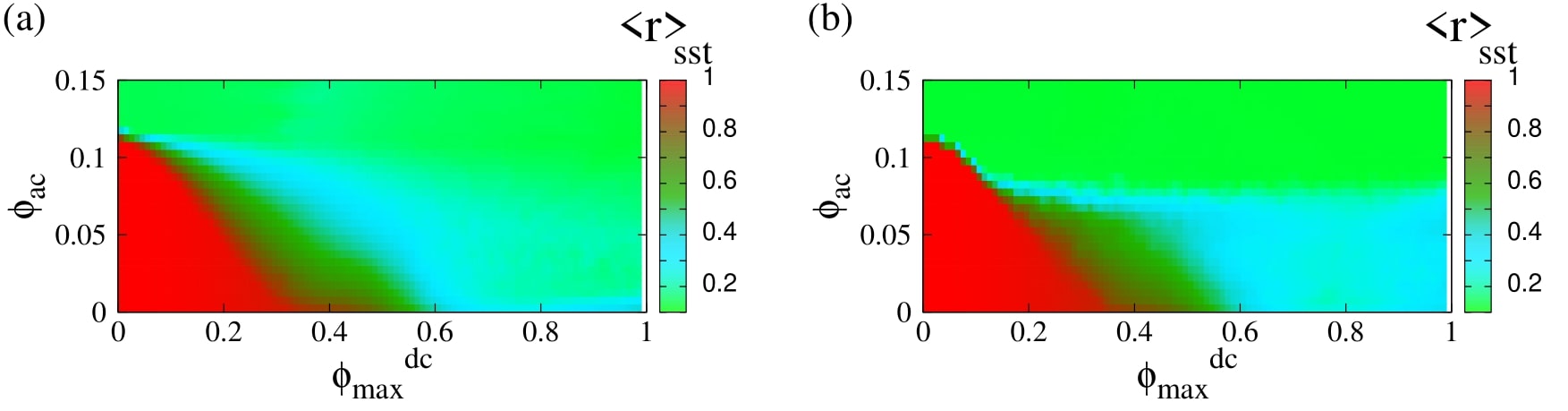} 
\end{center}
\caption{
 The magnitude of the synchronization parameter averaged over the steady-state 
 integration time ${\langle r \rangle}_{sst}$ mapped as a function of the ac flux 
 amplitude and the maximum dc flux bias ($\phi_{ac} - \phi_{max}^{dc}$ plane), 
 for $\beta_L =0.86$, $\gamma=0.01$, $N=54$, $\Omega =1.01$, and 
 (a) $\lambda=-0.02$, (b) $\lambda=-0.06$.
}
\label{fig13}
\end{figure}

\begin{figure}[h!]
\begin{center}
 \includegraphics[width=8.5cm]{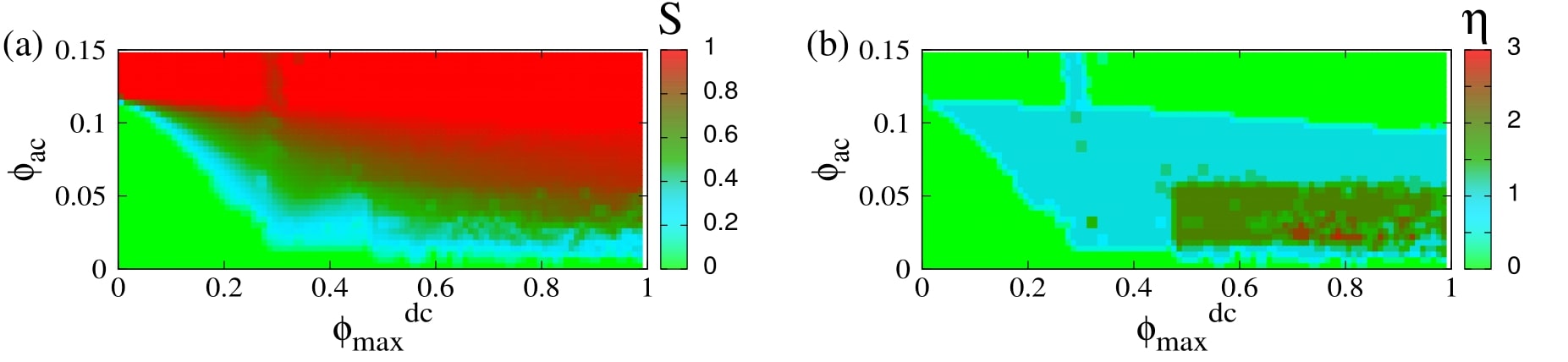} 
\end{center}
\caption{
 The index of incoherence $S$ and the chimera index $\eta$ are mapped on the 
 $\phi_{ac} - \phi_{max}^{dc}$ plane, for the same parameters as in Fig. 
 \ref{fig13}(a) and $n_0=4$, $\delta=10^{-4}$. 
}
\label{fig14}
\end{figure}

Two typical ``synchronization phase diagrams'', in which 
${\langle r \rangle}_{sst}$ is mapped on the $\phi_{ac} - \phi_{max}^{dc}$ 
parameter plane, are shown in Figs. \ref{fig13}(a) and (b) for $\lambda=-0.02$ 
and $\lambda=-0.06$, respectively. The frequency of the driving ac field has 
been chosen once again to be very close to the geometrical resonance of a single 
SQUID oscillator, i.~e., at $\Omega =1.01$. For each point on the 
$\phi_{ac} - \phi_{max}^{dc}$ plane, Eqs. (\ref{eq50}) are integrated in time 
with a standard fourth order Runge-Kutta algorithm using the initial conditions 
of Eq. (\ref{eq54}), with a time-step $h=0.02$. First, Eqs. (\ref{eq50}) are 
integrated for $10^5~T$ time-units to eliminate transients, and then they are 
integrated for $\tau_{sst} =10^5~T$ more time-units during which 
${\langle r \rangle}_{sst}$ is calculated. A comparison between Fig. \ref{fig13}(a) 
and (b) reveals that the increase of the coupling strength between 
nearest-neighboring SQUIDs from $\lambda=-0.02$ to $\lambda=-0.06$ results in 
relatively moderate, quantitative differences only. In both Figs. \ref{fig13}(a) 
and (b), for values of $\phi_{ac}$ and $\phi_{max}^{dc}$ in the red areas, the 
state of the SQUID metamaterial is synchronized. For values of $\phi_{ac}$ and 
$\phi_{max}^{dc}$ in the dark-green, light-green and light-blue areas, the state 
of the SQUID metamaterial is either completely desynchronized, or a chimera state 
with one or more desynchronized clusters. In order to obtain more information 
about these states, additional measures should be used, such as the incoherence 
index $S$ and the chimera index $\eta$ \cite{Gopal2014,Gopal2018}. These are 
defined as follows: First, define 
\begin{equation}
\label{eq55}
   v_n (\tau) \equiv {\langle \dot{\phi}_n \rangle}_T (\tau),
\end{equation}
where the angular brackets denote averaging over $T$, and 
\begin{equation}
\label{eq56}
   \bar{v}_n (\tau) \equiv \frac{1}{n_0 +1} \sum_{n=-n_0/2}^{+n_0/2} v_n (\tau),
\end{equation}
the local spatial average of $v_n (\tau)$ in a region of length $n_0+1$ around 
the site $n$ at time $\tau$ ($n_0 < N$ is an integer). Then, the local standard 
deviation of $v_n (\tau)$ is defined as 
\begin{equation}
\label{eq57}
   \sigma_n (\tau) \equiv \left< \sqrt{ \frac{1}{n_0 +1} \sum_{n=-n_0/2}^{+n_0/2} 
    \left( v_n -\bar{v}_n \right)^2 } \right>_{sst},
\end{equation}
where the large angular brackets denote averaging over the steady-state 
integration time. The {\em index of incoherence} is then defined as
\begin{equation}
\label{eq58}
   S=1 -\frac{1}{N} \sum_{n=1}^N s_n,
\end{equation}
where $s_n=\Theta(\delta -\sigma_n)$ with $\Theta$ being the Theta function. 
The index $S$ takes its values in $[0,1]$, with $0$ and $1$ corresponding to 
synchronized and desynchronized states, respectively, while all other values 
between them indicate the existence of a chimera or multi-chimera state. Finally, 
the {\em chimera index} is defined as 
\begin{equation}
\label{eq59}
   \eta =\sum_{n=1}^N |s_n -s_{n+1}| / 2,
\end{equation}
and takes positive integer values. The chimera index $\eta$ gives the number of 
desynchronized clusters of a (multi-)chimera state, except in the case of a 
completely desynchronized state where it gives zero. In Fig. \ref{fig14}, the 
incoherence index $S$ and the chimera index $\eta$ are mapped on the 
$\phi_{ac} - \phi_{max}^{dc}$ plane for the same parameters as in Fig. 
\ref{fig13}(a).
Figs. \ref{fig14}(a) and (b) provide more information about the state of the 
SQUID metamaterial at a particular point on the $\phi_{ac} - \phi_{max}^{dc}$
plane. In Fig. \ref{fig14}(a), for values of $\phi_{ac}$ and $\phi_{max}^{dc}$
in the light-green area ($S=0$) the SQUID metamaterial is in a synchronized 
state (see the corresponding area in Fig. \ref{fig14}(b) in which $\eta =0$). 
For values of $\phi_{ac}$ and $\phi_{max}^{dc}$ in the red area ($S=1$), the 
SQUID metamaterial is completely desynchronized (the corresponding area in 
Fig. \ref{fig14}(b) has $\eta =0$ due to technical reasons). For values of 
$\phi_{ac}$ and $\phi_{max}^{dc}$ in one of the other areas, the SQUID 
metamaterial is in a chimera state with one, two, or three desynchronized 
clusters, as it can be inferred from Fig. \ref{fig14}(b).    
\begin{figure}[h!]
\begin{center}
 \includegraphics[width=8.5cm]{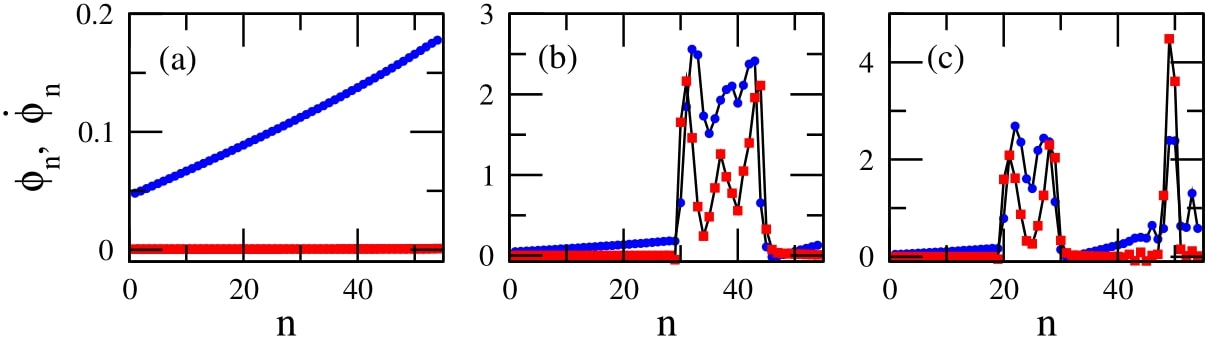} 
\end{center}
\caption{
 Flux and voltage profiles $\phi_n$ (blue) and $v_n =\dot{\phi}_n$ (red), 
 respectively, as a function of $n$ for $\beta_L =0.86$, $\gamma=0.01$,
 $\Omega =1.01$, $\phi_{ac} =0.04$, and (a) $\phi_{max}^{dc} =0.2$, 
 (b) $\phi_{max}^{dc} =0.4$, (c) $\phi_{max}^{dc} =0.6$.
}
\label{fig15}
\end{figure}

Using the combined information from Figs. \ref{fig13} and \ref{fig14}, the form 
of the steady-state of a SQUID metamaterial can be predicted for any physically 
relevant value of $\phi_{ac}$ and $\phi_{max}^{dc}$. In Fig. \ref{fig15}, three 
flux profiles $\phi_n$ are shown as a function of $n$, along with the 
corresponding profiles of their time-derivatives, $\dot{\phi}_n$. The profiles 
in Figs. \ref{fig15}(a), (b), and (c), are obtained for $\phi_{ac} =0.04$ 
and $\phi_{max}^{dc} =0.2$, $0.4$, and $0.6$, respectively, which are located in 
the light-green, light-blue, and dark-green area of Fig. \ref{fig15}(b). As it 
is expected, the state in Fig. \ref{fig15}(a) is an almost synchronized one, in 
Fig. \ref{fig15}(b) is a chimera state with one desynchronized cluster, while in 
Fig. \ref{fig15}(c) is a chimera state with two desynchronized clusters. At this 
point, the use of the expression ``almost synchronized'' should be explained. In 
the presence of a dc flux gradient, it is impossible for a SQUID metamaterial to 
reach a completely synchronized state. This is because each SQUID is subject
to a different dc flux, which modifies accordingly its resonance 
(eigen-)frequency. As a result, the flux oscillation amplitudes of the SQUIDs, 
whose oscillations are driven by the ac flux field of amplitude $\phi_{ac}$ and 
frequency $\Omega$, are slightly different. On the other hand, the maximum of 
the flux oscillations for all the SQUIDs is attained at the same time. Indeed, 
as can be observed in Fig. \ref{fig15}(a). the flux profile $\phi_n$ is not 
horizontal, as it should be in the case of complete synchronization. Instead,
that profile increases almost linearly from $n=1$ to $n=N$ (that increase is 
related to the dc flux gradient). However, the voltage profile $\dot{\phi}_n$ is 
zero for any $n$, indicating that all the SQUID oscillators are in phase. Since, 
in such a state of the SQUID metamaterial there is phase synchronization but no 
amplitude synchronization, the synchronization is not complete. However, the 
value of ${\langle r \rangle}_{sst}$ in such a state is in the worst case higher 
than $0.96$ for moderately high values of $\phi_{ac} =0.02 -0.10$ 
(Fig. \ref{fig12}(j)), which is a very high degree of global synchronization. 
Furthermore, the synchronized clusters in the chimera state profiles in Figs. 
\ref{fig15}(b) and (c), whose length coincides with that of the horizontal 
segments of the $\dot{\phi_n}$ profiles, also exhibit a very high degree of 
global synchronization (${\langle r \rangle}_{sst}> 0.96$). 

\section{Discussion and Conclusions}
The emergence of chimera and multi-chimera states in a 1D SQUID metamaterial 
driven by an ac flux field is demonstrated numerically, using a well-established 
model that relies on equivalent electrical circuits. Chimera states may emerge 
both with local coupling between SQUID (nearest-neighbor coupling) and nonlocal 
coupling between SQUIDs which falls-off as the inverse cube of their 
center-to-center distance. A large variety of initial conditions can generate 
chimera states which persist for very long times. In the previous Sections, the 
expression ``steady-state integration time'' is used repeatedly; however, in 
some cases this may not be very accurate, since chimera states are generally 
metastable and sudden changes may occur at any instant of time-integration
which results in sudden jumps the synchronization parameter ${\langle r \rangle}_T$
\cite{Lazarides2015b}. For the chimera states presented here, however, no such 
sudden changes have been observed.
Along with the ac flux field, a dc flux bias, 
the same at any SQUID, can be also applied to the 1D SQUID metamaterial. Chimera 
states can be generated in that case as well, although not shown here.

The emergence of those counter-intuitive states, their form and their global 
degree of synchronization depends crucially on the initial conditions. If the 
SQUID metamaterial is initialized with zeros, the generation of chimera states
does not seem to be possible for spatially constant dc flux bias $\phi_{dc}$.
In that case, synchronization-desynchronization and reverse 
synchronization-desynchronization transitions may occur by varying 
the ac flux amplitude $\phi_{ac}$ or the dc flux bias $\phi_{dc}$. In the former 
transition, a completely synchronized state suddenly becomes a completely 
desynchronized one. The replacement of the spatially constant dc flux bias by a 
position-dependent one, $\phi_n^{dc}$, makes possible the generation of chimera 
states from zero initial conditions. Here, a dc flux gradient is applied to the 
SQUID metamaterial, which provides the possibility to control the chimera state. 
Indeed, it is demonstrated that the position of the desynchronized cluster(s) 
and the global degree of synchronization can be controlled to some extent by 
varying the dc flux gradient. Moreover, in the presence of a dc flux gradient, 
the ac flux amplitude controls the size of the desynchronized cluster.

Here, the driving frequency is always chosen to be very close to the geometrical
frequency of the individual SQUIDs. In the case of relatively strong 
nonlinearity, considered here, the resonance frequency of individual SQUIDs is 
shifted to practically around the geometrical frequency. That is, for relatively 
strong nonlinearity, the driving frequency was chosen so that the SQUIDs are at 
resonance. For a single SQUID driven close to its resonance, the relatively 
strong nonlinearity makes it highly multistable; then, several stable and 
unstable single SQUID states may coexist (see the snake-like curves presented in 
Section 2). This dynamic multistability effect is of major importance for the 
emergence of chimera states in SQUID metamaterials, as it is explained below.

The dynamic complexity of $N$ SQUIDs which are coupled together increases with 
increasing $N$; this effect has been described in the past for certain arrays of 
coupled nonlinear oscillators as attractor crowding 
\cite{Wiesenfeld1989,Tsang1990}. This complexity is visible already for two 
coupled SQUIDs, where the number of stable states close to the geometrical 
resonance increases more than two times compared to that of a single SQUID
\cite{Hizanidis2016a}; some of these states can even be chaotic. Interestingly,
the existence of homoclinic chaos in a pair of coupled SQUIDs has been proved
by analytical means \cite{Agaoglou2015,Agaoglou2017}.
It has been argued that the number of stable limit cycles (i.~e., periodic 
solutions) in such systems scales with the number of oscillators $N$ as 
$(N-1)!$. As a result, their basins of attraction crowd more and more tightly 
in phase space with increasing $N$. The multistability of individual SQUIDs 
around the resonance frequency enhances the attractor crowding effect in SQUID 
metamaterial. Apart from the large number of periodic solutions (limit cycles), 
a number of coexisting chaotic solutions may also appear as in the two-SQUID 
system. All these states are available for each SQUID to occupy. Then, with 
appropriate initialization of the SQUID metamaterial, or by applying a dc flux 
gradient to it, a number of SQUIDs that belong to the same cluster may occupy 
a chaotic state. The flux oscillations of these SQUIDs then generally differ in 
both their amplitude and phase, resulting for that cluster to be desynchronized. 
Alternatingly, a number of SQUIDs that belong to the same cluster may find 
themselves in a region of phase-space with a high density of periodic solutions. 
Then, the flux in these SQUID oscillators may jump irregularly from one periodic 
state to another resulting in effectively random dynamics and in effect for that 
cluster to be desynchronized. At the same time, the other cluster(s) of SQUIDs 
can remain synchronized and, as a result, a chimera state emerges. 

\section*{Conflict of Interest Statement}
The authors declare that the research was conducted in the absence of any 
commercial or financial relationships that could be construed as a potential 
conflict of interest.

\section*{Author Contributions}
NL conceived the structure of the manuscript, JH and NL performed the
simulations, and all authors listed performed data analysis and have made 
intellectual contribution to the work, and approved it for publication.

\section*{Acknowledgments}
This research has been financially supported by the
General Secretariat for Research and Technology (GSRT) and the Hellenic 
Foundation for Research and Innovation (HFRI) (Code: 203).

\bibliographystyle{frontiersinHLTH&FPHY} 


\end{document}